\documentclass{emulateapj}
\usepackage[normalem]{ulem}
\usepackage{lscape}
\usepackage{color}
\usepackage{url}
\begin{document}


\title{The Impact of Stellar Multiplicity on Planetary Systems, I.: The Ruinous Influence of Close Binary Companions}

\author{
Adam L. Kraus\altaffilmark{1}, 
Michael J. Ireland\altaffilmark{2},
Daniel Huber\altaffilmark{3,4,5}
Andrew W. Mann\altaffilmark{1},
Trent J. Dupuy\altaffilmark{1}
}

\altaffiltext{1}{Department of Astronomy, The University of Texas at Austin, Austin, TX 78712, USA}
\altaffiltext{2}{Australia National University, Australia}
\altaffiltext{3}{Sydney Institute for Astronomy (SIfA), School of Physics, University of Sydney, NSW 2006, Australia}
\altaffiltext{4}{SETI Institute, 189 Bernardo Avenue, Mountain View, CA 94043, USA}
\altaffiltext{5}{Stellar Astrophysics Centre, Department of Physics and Astronomy, Aarhus University, Ny Munkegade 120, DK-8000 Aarhus C, Denmark}

\begin{abstract}

The dynamical influence of binary companions is expected to profoundly influence planetary systems. However, the difficulty of identifying planets in binary systems has left the magnitude of this effect uncertain; despite numerous theoretical hurdles to their formation and survival, at least some binary systems clearly host planets. We present high-resolution imaging of 382 Kepler Objects of Interest (KOIs) obtained using adaptive-optics imaging and nonredundant aperture-mask interferometry (NRM) on the Keck-II telescope. Among the full sample of 506 candidate binary companions to KOIs, we super-resolve some binary systems to projected separations of $<$5 AU, showing that planets might form in these dynamically active environments. However, the full distribution of projected separations for our planet-host sample more broadly reveals a deep paucity of binary companions at solar-system scales. For a field binary population, we should have found 58 binary companions with projected separation $\rho < 50$ AU and mass ratio $q > 0.4$; we instead only found 23 companions (a 4.6$\sigma$ deficit), many of which must be wider pairs that are only close in projection. When the binary population is parametrized with a semimajor axis cutoff $a_{cut}$ and a suppression factor inside that cutoff $S_{bin}$, we find with correlated uncertainties that inside $a_{cut} = 47^{+59}_{-23}$ AU, the planet occurrence rate in binary systems is only $S_{bin}=0.34^{+0.14}_{-0.15}$ times that of wider binaries or single stars. Our results demonstrate that a fifth of all solar-type stars in the Milky Way are disallowed from hosting planetary systems due to the influence of a binary companion.

\end{abstract}

\keywords{}

\section{Introduction}

\indent Radial velocity surveys and ground-based transit searches have discovered nearly 1000 confirmed planets around other stars \citep{Wright:2011kq} and thousands of additional candidates \citep{Borucki:2010fv,Batalha:2012oz,Burke:2014la}, revolutionizing the demography of planetary systems (e.g., \citealt{Fischer:2005gq,Johnson:2010kx,Bowler:2010rt}). The emerging consensus is that planetary systems are ubiquitious \citep{Dressing:2013kx,Petigura:2013ij,Foreman-Mackey:2014th,Muirhead:2015fy}, occurring with a frequency near unity across a wide range of stellar masses. However, most previous planet searches have only targeted single stars and very wide binaries since close companions complicate the observations and analysis. The majority of all solar-type stars form with at least one binary companion \citep{Duquennoy:1991zh,Raghavan:2010sz,Kraus:2008zr,Kraus:2011tg}, so the impact of stellar binarity on planet occurrence could represent one of the largest remaining systematic uncertainties in the Kepler era.

Binary companions should have a profound dynamical influence on the planet formation process, truncating disks \citep{Artymowicz:1994ir,Jang-Condell:2008qa,Andrews:2010ec,Jang-Condell:2015rw}, dynamically stirring planetesimals \citep{Quintana:2007lr,Haghighipour:2007vn,Rafikov:2015pb,Silsbee:2015dk}, and enhancing both accretion \citep{Artymowicz:1994ir,Jensen:2007zo} and photoevaporation \citep{Alexander:2012la}. These processes suggest that disks in binary systems could be short-lived and that such systems would represent hostile sites for planets to grow. Furthermore, even if planets can be formed, then secular evolution of the orbits and even stellar evolution can drive systems through unstable states that destroy them on Myr to Gyr timescales \citep{Holman:1999yu,Haghighipour:2006qf,Kratter:2012kl,Kaib:2013yq}, though dynamical interactions also have been suggested as a channel for producing the many giant planets discovered well inside the snow line (\citealt{Kozai:1962vn,Fabrycky:2007rt,Winn:2010ys,Naoz:2012rr}; but also \citealt{Ngo:2015fr}).

In spite of the dynamical barriers, ground-based exoplanet surveys show that some binary systems do host planets. A handful of giant planets have been identified in nearby short-period binary systems \citep{Hatzes:2003rt,Correia:2008ai,Kane:2015fk}, though it has been suggested that they might result from small-N dynamical interactions rather than in-situ formation (e.g. Pfahl \& Muterspaugh 2006). Circumbinary gas giants are now being reported by Kepler \citep{Doyle:2011qf}, perhaps as frequently as for single stars \citep{Welsh:2012lr}, and orbit monitoring for planet-hosting binary KOIs is also uncovering a population of small planets in close binaries (e.g., Kepler-444; Dupuy et al. 2016). Numerous wide binary companions to planet hosts have been identified as well \citep{Mugrauer:2005zr,Daemgen:2009my,Muterspaugh:2010mz,Bergfors:2012qe}, and it appears that wide binary systems are common exoplanet hosts \citep{Eggenberger:2007vn,Desidera:2007yq,Duchene:2010cj}. However, the observational biases against exoplanet discovery in the presence of a companion star mean that the frequency, properties, and provenance of planets in close binary systems are still largely unconstrained by data; some ground-based RV surveys have been launched and have yet to report any detections (e.g., Eggenberger 2010; Desidera et al. 2010), but the final statistics have not been reported yet. These inputs will be crucial as theory investigates the circumstances under which planets apparently can form in binary systems (e.g., \citealt{Jang-Condell:2008qa,Rafikov:2013fu,Jang-Condell:2015rw,Rafikov:2015dz}).

This question has long been considered from a planet formation perspective, where the occurrence and properties of protoplanetary disks can be compared for binary systems versus single stars \citep{Ghez:1997aa,White:2001jf,Cieza:2009fr,Duchene:2010cj} to demonstrate a striking depletion of protoplanetary disks among close binary systems. This trend emerged most clearly from a combination of a binary census of the Taurus-Auriga star-forming region with a disk census from Spitzer \citep{Kraus:2011tg,Kraus:2012qe}, which showed that at the age of $\sim$2 Myr, 80\% of single stars and 90\% of wide binaries host a protoplanetary disk, whereas only 35\% of close ($<$40 AU) binaries host disks. An analysis of younger regions shows that this depletion occurs within $<$1 Myr \citep{Cheetham:2015qy}. The detailed properties of protoplanetary disks reveal an even more striking trend. \citet{Harris:2012oz} showed that even when binaries host protoplanetary disks, then disk masses are depleted by a factor of $\sim$5 for 30--300 AU binaries and by a factor of $\sim$25 for $<$30 AU binaries; the only exception is a small number of circumbinary disks which are quite massive. However, detailed spectroscopic studies of many of the same protoplanetary disks \citep{Pascucci:2008uq,Skemer:2011ij} demonstrate that the surviving disks do undergo the same evolutionary processes of grain growth and dust settling. Studies of intermediate-age debris disk hosts also reveal that debris exists in binary systems (Trilling et al. 2007; Rodriguez et al. 2015), albeit suppressed by a factor of $\sim$3 for 1--50 AU systems when compared to tighter or wider systems.

The recent explosion in exoplanet discoveries from the Kepler mission (Borucki et al. 2010) offers a new opportunity to characterize the influence of stellar multiplicity on planet occurrence. The coarse spatial resolution and nearly blind target selection of Kepler with respect to stellar multiplicity have rendered the Kepler Objects of Interest (KOIs; \citealt{Batalha:2012oz}) largely unbiased to stellar multiplicity, offering the first planet sample for which the presence of close binaries is uncorrelated with planet discovery. Numerous high-resolution imaging surveys have targeted Kepler planet hosts \citep{Adams:2012zr,Adams:2013ve,Lillo-Box:2012uq,Lillo-Box:2014ul,Law:2013fk,Dressing:2014qf,Wang:2014uq}, discovering a large number of wide ($\ga$100 AU) binary companions and vetting many of the KOIs to eliminate blends and other sources of false positives. The sum of those observations seem to suggest that the wide binary population is in line with that seen for the full stellar population \citep{Horch:2014pd,Deacon:2016lr}, but since KOIs are distant (in all but a handful of cases, $d \ga 100$ pc), then conventional AO imaging can not access binary companions on solar-system scales ($\rho \sim$ 5--20 AU). RV monitoring has indicated at $\sim$2$\sigma$ that the close ($\la 20$ AU) binary occurrence rate might be suppressed among the set of KOIs with multiple observations (e.g., Wang et al. 2015a), but with uncertain selection effects resulting from the choice of which targets merit RV followup.

In this paper, we report nonredundant aperture-mask interferometry and adaptive optics imaging of 382 KOIs at higher spatial resolution, capable of detecting binary companions down to Solar System scales (inner working angles of 2--5 AU). We find a striking paucity of close companions to planet host stars, which supports theoretical predictions that these binary systems (which are known to represent $\sim$25\% of all Sun-like stars) must be extremely hostile sites for extrasolar planets to form and/or survive. In Section 2, we describe our selection of a volume-limited sample of Kepler planet hosts, and in Section 3, we describe our observations and data analysis. In Section 4, we describe the binary population unveiled by our survey and update the properties of the planet hosts and planetary systems in light of the flux contributions of the previously-unidentified binary companions. Finally, in Section 5, we discuss the properties of the binary population among planet host stars, and infer the impact of those binary companions on the formation and survival of extrasolar planetary systems.

\section{The Sample}

 \begin{figure*}
 \epsscale{0.95}
 \plotone{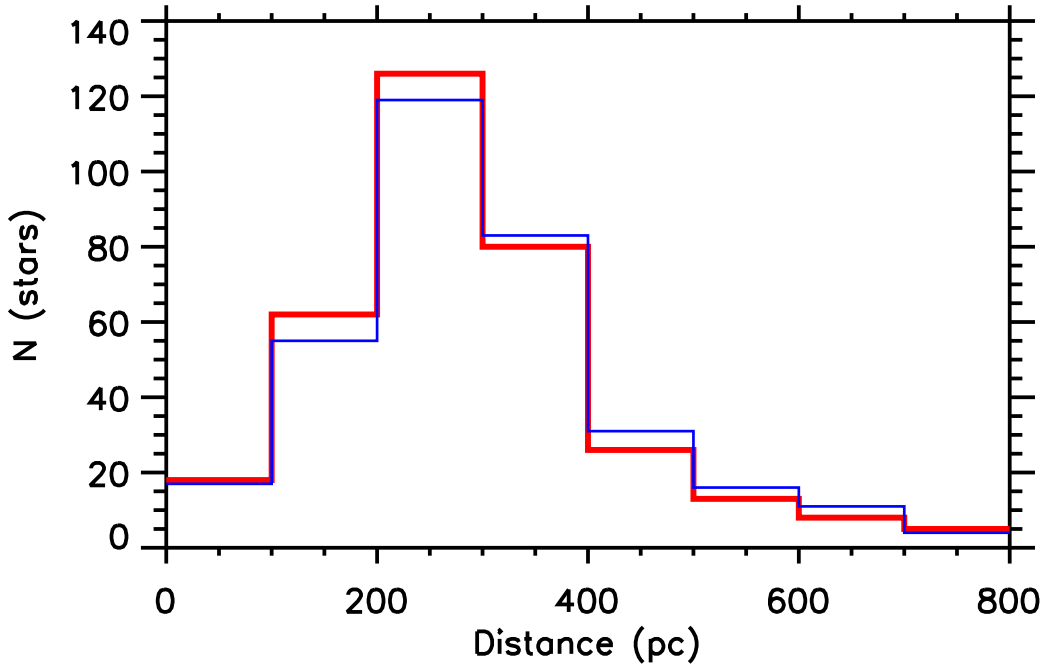}
 \caption{Distance distribution for our observed sample, where red shows the initial distance estimates and blue shows distances corrected for flux contributions from binaries (Malmquist bias). To maintain clarity for the majority of the sample, the extended tail of 19 targets with $d > 800$ pc (and extending to $>$2000 pc, for some giants) are not shown. If the population were distributed uniformly in space, there should be 2.4 times as many targets at $200 < d < 300$ pc as at $d < 200$, indicating that the flux-limited sample of KOIs is incomplete beyond 200 pc. However, by weighting binaries with a $1/V_{max}$ weight, we can still control for the Malmquist bias introduced by the $K_p \sim$16 limit of Kepler and by the enhanced planet detectability seen for bright stars with high-SNR Kepler data.}
  \end{figure*}

 \begin{figure}
 \epsscale{1.15}
 \plotone{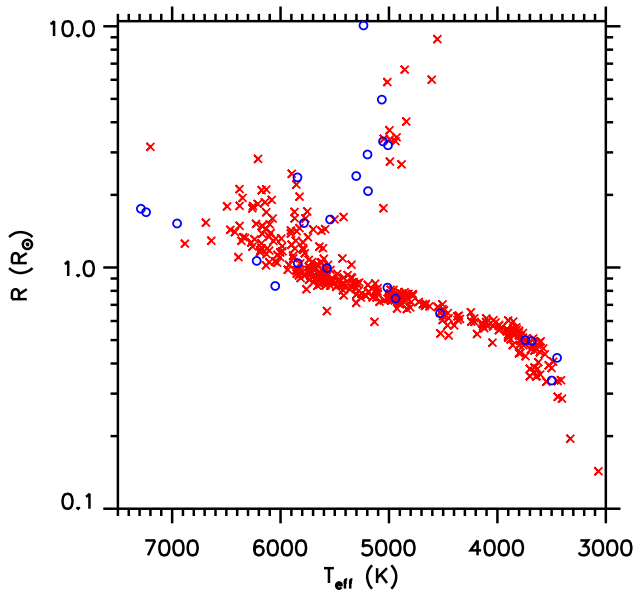}
 \caption{HR diagram (radius versus temperature) for our observed sample. Candidate and confirmed planet hosts are shown with red crosses, while false positives that we omit from our statistical analysis are shown with blue open circles. The vast majority of targets are main sequence stars. Our observations of giants are less likely to detect low-mass companions because the large distance imposes a penalty on both sensitivity and resolution. However, our detection limits account for the distance in turning observational limits (in magnitudes and arcseconds) into physical limits (in mass ratios and AU), so we retain the giants in order to maintain a uniform treatment of all stars.}
  \end{figure}

The Kepler target list \citep{Batalha:2010lr} consists of $2 \times 10^5$ stars which are optically bright ($K_p \la 16$) and comprises $10^5$ G dwarfs (explicitly selected to be suitable for detecting Earth analogs), $2 \times 10^4$ F dwarfs, $2 \times 10^4$ K dwarfs, $3 \times 10^3$ M dwarfs, $10^3$ OBA dwarfs, and at least $1.5 \times 10^4$ giants of all types \citep[e.g.][]{hekker11,Mann:2012ul,stello13}. The Kepler targets were selected from the Kepler Input Catalog, which was constructed using broadband ($grizJHK_s$) and intermediate-band (D51) photometry to infer $T_{eff}$, $\log(g)$, [Fe/H], and $A_V$ for $1.3 \times 10^7$ sources in the Kepler field. As of 2013-07-31, the Kepler survey had yielded 4914 Kepler Objects of Interest (KOIs) that included 3548 candidate or confirmed planets transiting 2664 host stars.\footnote{\url{http://exoplanetarchive.ipac.caltech.edu/}} For this work, we use the KOI list from that epoch as an input sample.

Most of our targets represent a volume-based subsample of KOIs, selected from the July 2013 list using spectrophotometric distances. We initially estimated the effective temperature $T_{eff}$ and the bolometric flux $m_{bol}$ based on two methods. First, we used SED fitting of photometry from 2MASS \citep{Skrutskie:2006nb}, SDSS \citep{Ahn:2012ys}, and USNO-B1.0 \citep{Monet:2003yt}, as described in \citet{Kraus:2007mz} and \citet{Kraus:2014rt}, to estimate both parameters. Second, we directly adopted the KIC $T_{eff}$, and then combined a relation between $K_p$ and $g'r'$ in the Kepler documentation\footnote{\url{http://keplergo.arc.nasa.gov/CalibrationZeropoint.shtml}} with the \citet{Kraus:2007mz} relation between $g$/$r$/$T_{eff}$ and $m_{bol}$ to derive a relation between $K_p$ and $m_{bol}$. In both cases, we then used the \citet{Kraus:2007mz} relation between $T_{eff}$ and $M_{bol}$ to compute the photometric distance modulus, $DM = m_{bol} - M_{bol}(T_{eff})$, and then prioritized targets in order of ascending distance.

After the majority of these observations had occurred, we adopted updated distances based on the latest version of the Kepler stellar properties catalog (Q1-17; DR24\footnote{\url{http://exoplanetarchive.ipac.caltech.edu/docs/Kepler\_stellar\_docs.html}}), initially described by \citet{Huber:2014uo}. To derive self-consistent stellar properties and distances, we followed the method by \citet{serenelli13} to calculate posterior distribution functions for each model parameter by marginalizing isochrones from the Dartmouth Stellar Evolution Program \citep{Dotter:2008qq} conditioned on literature values for atmospheric properties ($T_{eff}$, $\log(g)$, and $[Fe/H]$). Distance posteriors were calculated using absolute K-band magnitudes from the isochrone grid and the reddening map by \citet{amores05}, assuming the KOI is a single star with the given stellar parameters. All reported stellar properties are based on the posterior mode, and uncertainties are calculated based on the closest $1\,\sigma$ interval around the mode.

Input values and uncertainties for $T_{eff}$, $\log(g)$, and $[Fe/H]$ were taken from the Kepler stellar properties catalog except for 13 M dwarfs for which we adopted new temperatures and metallicities based on medium resolution spectra obtained with the SNIFS spectrograph following the method given in \citet{Mann:2013fj}, and Kepler-444 for which we adopted the Hipparcos distance and input parameters by \citet{campante15}. In total, 74 stars (19\%) in our sample have input parameters from asteroseismology, 235 stars (62\%) from spectroscopy, and 73 stars (19\%) from broadband colors. By construction, our stellar properties are in close agreement with \citet{Huber:2014uo}, except for 21 K stars with poorly constrained photometric $\log(g)$ values which are classified as evolved ($\sim 2-4 R_{\odot}$) subgiants in \citet{Huber:2014uo}, but have main-sequence radii ($\sim 1 R_{\odot}$) in our work. This difference is due to the fact that the \citet{Huber:2014uo} estimates are based on maximum likelihood values while probabilistically, given the large uncertainty of 0.4\,dex in $\log(g)$, smaller radii are favored due to longer main-sequence lifetimes compared to the subgiant branch. Within the uncertainties, however, all classifications are consistent within 3\,$\sigma$.
 
The revised distances derived using the above procedure were used to rerank our list. Future observations will prioritize the closest KOIs from this list, and we hereafter use these updated distances (modified for the flux contributions of binary companions if necessary) to convert angular separations into physical separations. Since binaries are overrepresented in a flux-limited sample, our model comparisons will build in Malmquist bias with a $V/V_{max}$ weighting (e.g., Schmidt 1968) such that binaries with a flux addition from a companion are predicted with greater frequency set by the fractionally larger volume within which they occur. We also report observations for 23 stars that were subsequently identified as likely false positives (e.g., \citealt{Slawson:2011ec,Mann:2012ul,Huber:2013ly}) after we observed them, since high-resolution imaging is useful for further confirming their non-planetary nature, but do not use them in the statistical analysis.

We list all of the observed KOIs and their stellar parameters in Table 1, divided into the likely planet hosts (359 KOIs that have been confirmed or remain as candidates) and the false positives (23 stars that have been rejected). Where available, we list Kepler numbers for the hosts of confirmed planets and the reference by which the false positives have been rejected. Many of the false positives are labeled as such on the Kepler Community Follow-On Project (CFOP) website\footnote{\url{https://exofop.ipac.caltech.edu/cfop.php}}, so we attribute those labels to specific community members when their identities are known. In Figure 1, we present a distance histogram for the statistical sample. In Figure 2, we present an HR diagram for the sample using the stellar parameters reported by \citet{Huber:2014uo} and our inferred distances. The final sample of 382 stars consists of 3 A stars, 52 F stars, 137 G stars, 134 K stars, and 56 M stars, based on the $T_{eff}$-SpT relation of \citet{Kraus:2007mz}. The distribution of AFG stars is roughly consistent with the present-day mass function of the solar neighborhood (e.g., \citealt{Reid:2002sn}), but there is a distinct paucity of K and especially M stars due to the emphasis on solar analogs in the Kepler target list.

\section{Observations and Data Analysis}

\subsection{High-Resolution Imaging and Nonredundant Mask Interferometry Observations}

The technique of non-redundant aperture masking (NRM) has been well established as a means of achieving the full diffraction limit of a single telescope \citep{Nakajima:1989zl,Tuthill:2000ge,Tuthill:2006kq,Ireland:2013lr}, beyond what can be achieved with standard AO imaging. The core innovation of aperture masking is to resample the telescope's single aperture into a sparse interferometric array; this allows for data analysis using interferometric techniques (such as closure-phase analysis) that calibrate out the phase errors that limit traditional astronomical imaging by inducing speckle noise. As we described in \citet{Kraus:2008zr,Kraus:2011tg}, aperture-masking observations can yield contrasts as deep as $\Delta K \sim 6$ at $\lambda / D$ and $\Delta K \sim 4$ at 1/3 $\lambda /D$ with observations of $\sim$5--15 minutes, and we have used the technique to identify dozens of binary companions that fall inside the detection limits of traditional imaging surveys. More detailed discussions of the benefits and limitations of aperture masking, as well as typical observing strategies, can be found in \citet{Kraus:2008zr} and in \citet{Readhead:1988so,Nakajima:1989zl,Tuthill:2000ge,Tuthill:2006kq,Lloyd:2006kl,Martinache:2007ru,Ireland:2008yq}.

We observed our targets with the Keck-II telescope and either natural guide star (NGS) or laser guide star (LGS) AO in vertical angle mode. Our observations were taken over the space of 22 half or full nights between 2012 May and 2014 August. All observations were conducted with the facility adaptive optics imager NIRC2, which also has a 9-hole aperture mask installed in a cold filter wheel near the pupil stop. All observations used the smallest pixel scale ($9.952 \pm 0.002$ mas/pix; \citealt{Yelda:2010qf}) and we corrected for geometric distortion using their NIRC2 distortion solution. Each observing sequence consisted of four steps:

\begin{enumerate}

\item AO acquisition, requiring 1 minute for NGSAO or 4 minutes for LGSAO.

\item Shallow imaging, requiring 1 minute to obtain 1--2 integrations of 10--20 seconds with few Fowler samples. These observations were intended for initial target acquisition and to reconnoiter for obvious binarity, as well as offering sensitivity to some additional companions between the inner working angle of the coronagraph ($\rho \sim 400$ mas) and the typical angle inside of which NRM supercedes imaging ($\rho \sim 150$ mas).

\item Deep imaging, requiring 2 minutes to obtain 2 integrations of 20 seconds with many Fowler samples. These observations were intended to search for faint companions at wide projected separations ($\rho > 0.5\arcsec$). Targets brighter than $K_{2M} \sim 10.6$ were placed behind the 600 mas coronagraphic spot to avoid saturation of the primary, since many Fowler samples can only be conducted with long integrations. Fainter targets were observed without the coronagraph because there was insufficient flux for the primary to be detected through the coronagraph (impeding the measurement of precise astrometry for candidate companions).

\item Nonredundant mask interferometry, requiring 3--4 minutes to obtain 6--8 integrations of 20s with the 9H mask in place. We chose the number of observations to match target brightness and observing conditions, with the goal of achieving a contrast of $\Delta K' > 3$ at $\lambda /D$ even for the faintest targets and well above median seeing. The median contrast limit at the diffraction limit was $\Delta K' \sim 4.3$, while for bright targets in good conditions, the contrast limit typically was $\Delta K' \sim 5$. 

\end{enumerate}

The observing sequence was fully scripted, requiring $\sim$7 minutes for NGSAO observations and $\sim$10 minutes for LGSAO observations. A typical interferometric measurement requires the observation of one or more source-calibrator pairs. However, our sample included numerous targets with similar positions and brightnesses, so we instead observed groups of science targets and inter-calibrated between them, omitting binary systems from the calibration as they were identified. The actual number of observations obtained at each step can vary somewhat, due to ongoing optimization of our observing strategy, the rejection of bad frames (i.e., when the AO system lost lock), and some cases where we exited the script to obtain additional frames in compensation for obviously bad data (such as from bad seeing, windshake, or intermittent clouds).

We summarize the salient details of our imaging observations in the same tables as our detection limits (Table 2 and Table 4), as described below.

\subsection{Imaging Analysis for Isolated Primaries}

Each science frame was linearized using the IDL task \textit{linearize\_nirc2.pro}\footnote{\url{http://www.astro.subysb.edu/metchev/ao.html}}, and then dark-subtracted and flat-fielded using the most contemporaneous darks and flats from each run. We then identified dead pixels from an analysis of $K'$ superflats (constructed from at least 20 frames in one night) from 41 separate nights, spanning 2006-2013, identifying and interpolating over any pixel with a response of $<$30\% in at least half of all superflats. Similarly, we identified hot pixels from an analysis of long-integration superdarks (constructed from at least 30 frames in one night, with 1 coadd of $t_{int}=20$ sec, taken with 16 or more Fowler samples) from 38 separate nights, identifying any pixel as hot if it had $\ge$10 counts in at least half of the superdarks. Finally, we corrected cosmic rays and transient hot pixels by identifying all pixels with fluxes $> 10\sigma$ above the median of the 8 adjacent pixels or 16 once-removed pixels, replacing them with that median value.

For each science frame, we used two different methods of PSF subtraction. The first method of PSF subtraction, intended to find faint sources at wide separations (in the background- or readnoise-limited regime), was to subtract an azimuthal median PSF model. This PSF model was constructed from a five-pixel boxcar median, calculated in concentric rings around the science target and interpolated to the exact distance for every pixel. This method leaves all high-frequency structure (i.e., speckles and diffraction spikes) from the primary star PSF, so it is not ideal for identifying close ($\rho \la 1$\arcsec) companions. However, it has the virtue of adding negligible noise at large separations (as would occur by subtracting empirical PSF templates), and hence allowing for the most robust identification of wide candidate companions. Many of our observations had modest strehl (15--30\%), with much of the flux in a broad seeing-limited halo, so the subtraction of this smooth halo uncovered faint sources that otherwise would have been missed.

The second method of PSF subtraction, intended to find sources within the primary star's PSF halo, uses empirical PSFs of other targets (observed within the same run) to more closely match the primary star PSF for subtraction. Our observations are short and do not show substantial sky rotation, so for each science frame, we begin with an initial library of all science frames for other targets that were taken in the same filter and coronagraph (or lack thereof) and that have not been identified as binary systems. We then rescale each potential template frame to the science frame in question and measure the $\chi^2$ of the residuals in an annulus of $\rho = 150$--300 mas (for non-coronagraphic data) or $\rho = 450$--750 mas (for coronagraphic data). Since our imaging observations are not deep, the vast majority of structure in the PSF is represented by the diffraction spikes and by a few well-established, long-term superspeckles, and hence the number of free parameters in this fit is small. We therefore only use a single frame as a PSF template and globally optimize the fit, since otherwise the flux from any companions rapidly becomes a dominant source of residuals and therefore is fit and subtracted as well.

Using the coadded stacks of all median-subtracted and all template-subtracted frames on a science target, we identified candidate companions by measuring the flux through 80 mas (diameter) apertures centered on every pixel of the image. We then measured the standard deviation of the fluxes among all apertures in a sliding 5-pixel annulus around the primary star, identifying any aperture with a $> +6 \sigma$ outlier as the location of a potential astrophysical source. We set the corresponding detection limit for that projected separation to the contrast associated with that $+6 \sigma$ flux value (as measured in comparison to an identical aperture centered on the primary star). To reconcile the possible detections in the median-subtracted and template-subtracted images, we only accepted a potential astrophysical source if it was identified in both sets of residuals, or if it was identified in one set and fell below the detection limits for the other set. Hence, faint potential sources at wide projected separations typically would be included if they were in the median-subtracted image and not the template-subtracted image (but not vice versa), because the median-subtracted image had deeper limits. At small projected separations, the converse was true, because template-subtracted images had deeper limits. This process still passed through residual cosmic rays that had not escaped our early correction steps, as well as some faint artifacts along diffraction spikes and near the largest superspeckles, so we then visually inspected each remaining candidate to determine if it was a cosmic ray or corresponded to PSF features that could be seen in contemporaneous observations of other science targets.

Once a potential source was accepted as a bona fide astrophysical object (and hence a candidate companion), we measured aperture photometry for the candidate companion (in both the median-subtracted or template-subtracted images) and the science target (in the processed, unsubtracted images) in order to determine the relative astrometry and photometry of the candidate. By default, this aperture photometry used an aperture with diameter of 80 mas and a sky annulus with inner and outer radii of 100--150 mas. We visually inspected each companion to determine which PSF-subtraction technique produced a cleaner detection, and adopted that measurement for all subsequent analysis.

\subsection{Imaging Analysis for Close Pairs}

We found that our default imaging pipeline failed in cases where a candidate companion was bright enough to have a substantial PSF halo and that halo impinged significantly on the primary star, as it was no longer valid to fit the flux distribution with a single PSF template. We therefore also used an alternate pipeline for the production of PSF-subtracted images of multiple sources, which iteratively uses the best-fit PSF template to fit for the parameters of the binary or triple (projected separation, PA, and contrast of each other source with respect to the brightest object), and then creates double-star or triple-star templates of all empirical PSFs and tests them to find which empirical PSF is best. This process then iterated until the same template is used to find the same binary or trinary parameters, at which point the multi-PSF template is subtracted and the template-subtracted image is fed to subsequent pipeline steps for identification of further candidate companions. The PSF-fit values for projected separation, PA, and contrast are adopted in place of the aperture photometry described above.

\subsection{NRM Analysis}

The data analysis follows almost the same prescription as in \citet{Kraus:2008zr,Kraus:2011tg}, so we discuss here only a general background to the technique and differences from \citet{Kraus:2008zr}. The data analysis takes three broad steps: basic image analysis (flat-fielding, bad pixel removal, dark subtraction), extraction and calibration of squared visibility and closure phase, and binary model fitting. Unless fitting to close, near-equal binaries, we fit only to closure phase, as this is the quantity most robust to changes in the AO point-spread function (PSF). 

In previous papers we used Monte-Carlo simulations based on carefully modeled data covariance matrices in order to compute detection limits. In this work, the empirical closure-phase covariance matrix was not always easy to estimate, because the number of contemporaneous point sources observed with any target was heterogeneous with such a large data set. Instead, we chose a more conventional approach of scaling the uncertainties so that the best fit binary model to closure-phase had a reduced chi-squared of 1.0. Any companion solutions that had a significance of more than $6 \sigma$ were called detections, and non-detections were assigned a detection-limit equal to this $6 \sigma$ threshold; the detection limit reported is then the azimuthal average of this threshold. This limit was similar to (but in most cases a little more conservative than) the Monte-Carlo technique used in previous papers. As visibility amplitude is very useful for the closest binary solutions in breaking a contrast/separation degeneracy, we included visibility amplitude in our fits whenever the best fit solution using amplitudes had $\rho < 40$ mas and $\Delta K' < 1$ mag, and where the inclusion of visibility amplitudes did not reduce the significance below the $6 \sigma$ cutoff (as could happen in the case of poorly calibrated visibility amplitudes).

\subsection{Candidate Companion Stellar Properties}

Candidate companion properties were estimated in the same procedure as described in Section 2 for the primary. For each primary model of fixed age and metallicity, we used the observed flux ratio $\Delta K'$ to interpolate stellar properties to the absolute K magnitude of the secondary, and assigned the interpolated models the posterior probability of the corresponding primary model (thereby assuming the same age, metallicity and distance for both stars). Stellar properties were then derived by marginalizing the posterior probabilities and calculating the mode of the posterior distribution function for each stellar parameter. Uncertainties in $\Delta K'$ were taken into account by repeating the procedure for the $\pm 1\sigma$ limits in $\Delta K'$ and adding the resulting difference in quadrature to the estimate obtained with the observed $\Delta K'$ value.

In addition to secondary temperature, radius and mass, we additionally directly computed the mass ratio so that the correlated errors between secondary and primary mass (which results from the uncertain distance, age, and metallicity) partially cancel. Finally, we also use absolute Kepler bandpass magnitudes provided in the Dartmouth grid to estimate the flux ratio in the Kepler bandpass ($\Delta K_p$), which will be useful in removing flux dilution of transits and fitting updated planetary radii.

For all candidate companions, we estimate the projected separation (in AU) from the projected angular separation using the distances calculated in Section 2, after accounting for the excess flux due to the candidate companion(s). The original 2MASS photometry was conducted with an aperture of radius 4\arcsec, so we only include companions closer than that limit. The updated distances are reported in a separate column of Table 1.

 \begin{figure*}
 \epsscale{1.2}
 \plotone{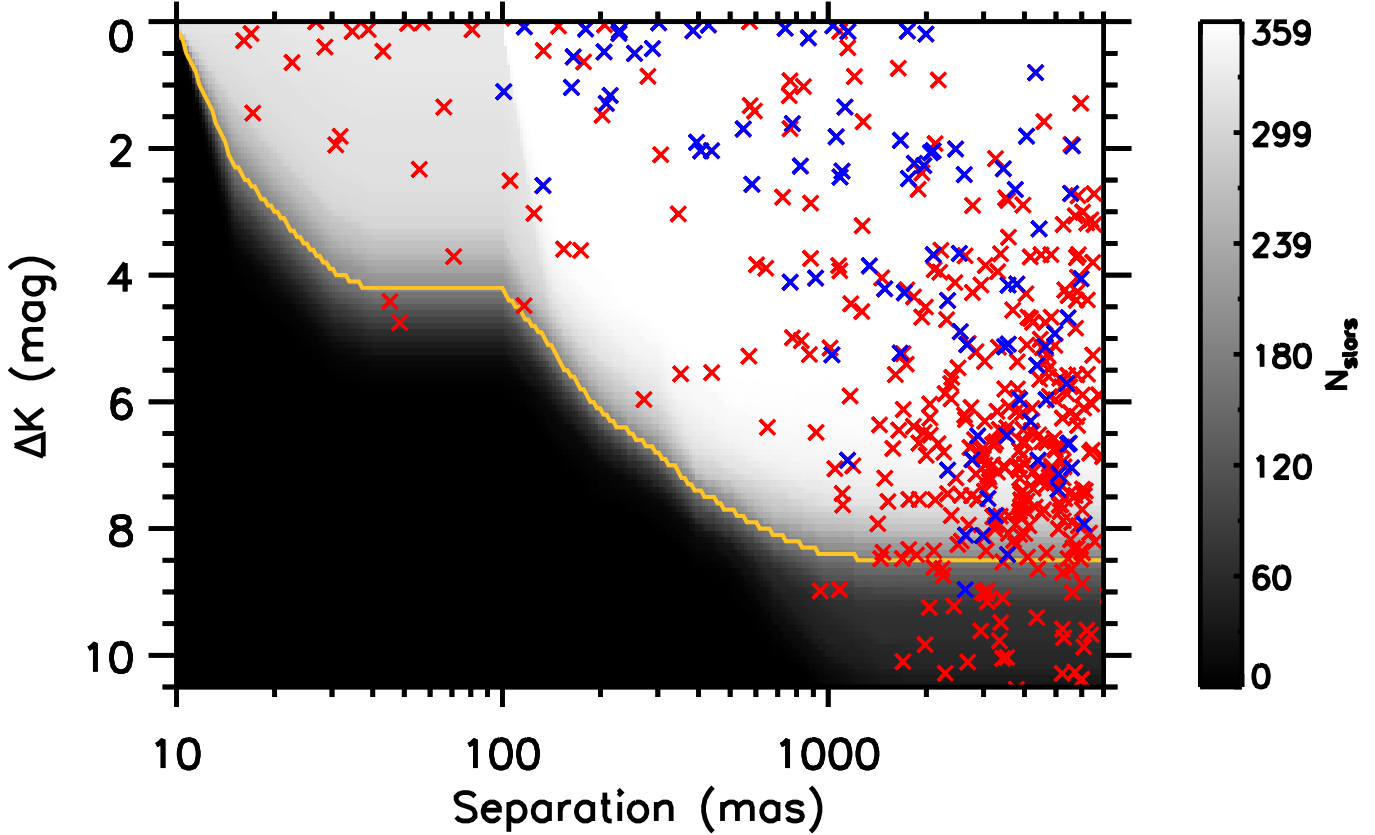}
 \caption{Detections and detection limits for our survey, in terms of contrast $\Delta K'$ in magnitudes and angular separation $\rho$ in milliarcseconds. The detections are shown with red crosses for newly detected sources and blue crosses for previously identified sources, while the detection limits are shown in a shaded background trending from black (no observations sensitive to that combination of contrast and separation) to white (all 359 planet hosts that are sensitive to that combination). The orange solid line shows the median limit for the survey. There were no known binaries among this volume-limited sample with $\rho < 0.1$\arcsec, the regime uniquely probed by NRM (upper left) that represents the potential for discovering binary companions on solar system scales.}
  \end{figure*}

 \begin{figure*}
 \epsscale{1.2}
 \plotone{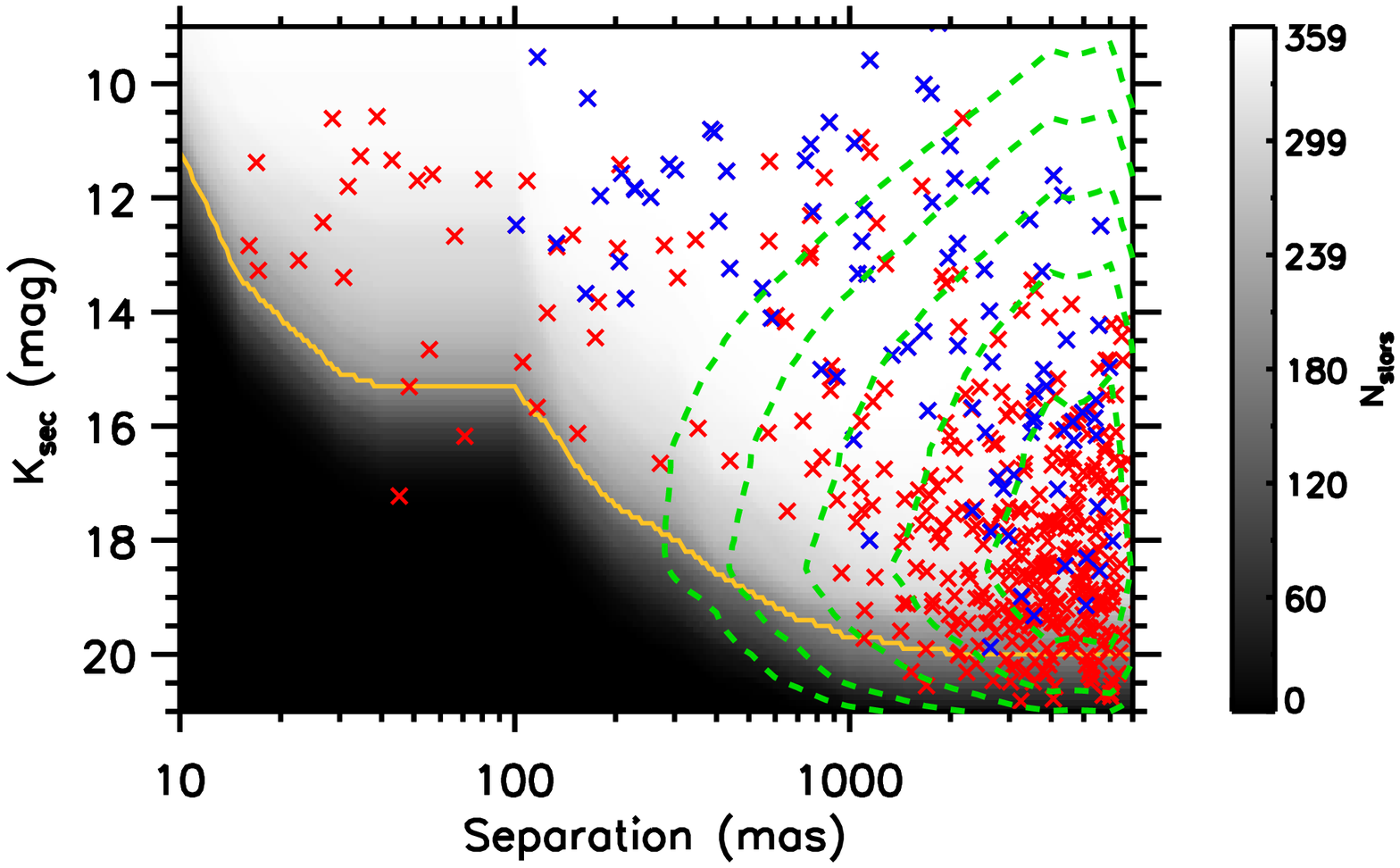}
 \caption{Detections and detection limits for our survey, in terms of companion apparent magnitude $K'$ in magnitudes and angular separation $\rho$ in milliarcseconds. The detections are shown with red crosses for newly detected sources and blue crosses for previously identified sources, while the detection limits are shown in a shaded background trending from black (no observations sensitive to that combination of contrast and separation) to white (all 359 planet hosts that are sensitive to that combination). The orange solid line shows the median limit for the survey. The green dashed lines show the expected contours of the field contaminant distribution, drawn such that 1, 3, 10, 30, and 100 companions should fall outside of the concentric contours.}
  \end{figure*}
  
 \begin{figure*}
 \epsscale{1.2}
 \plotone{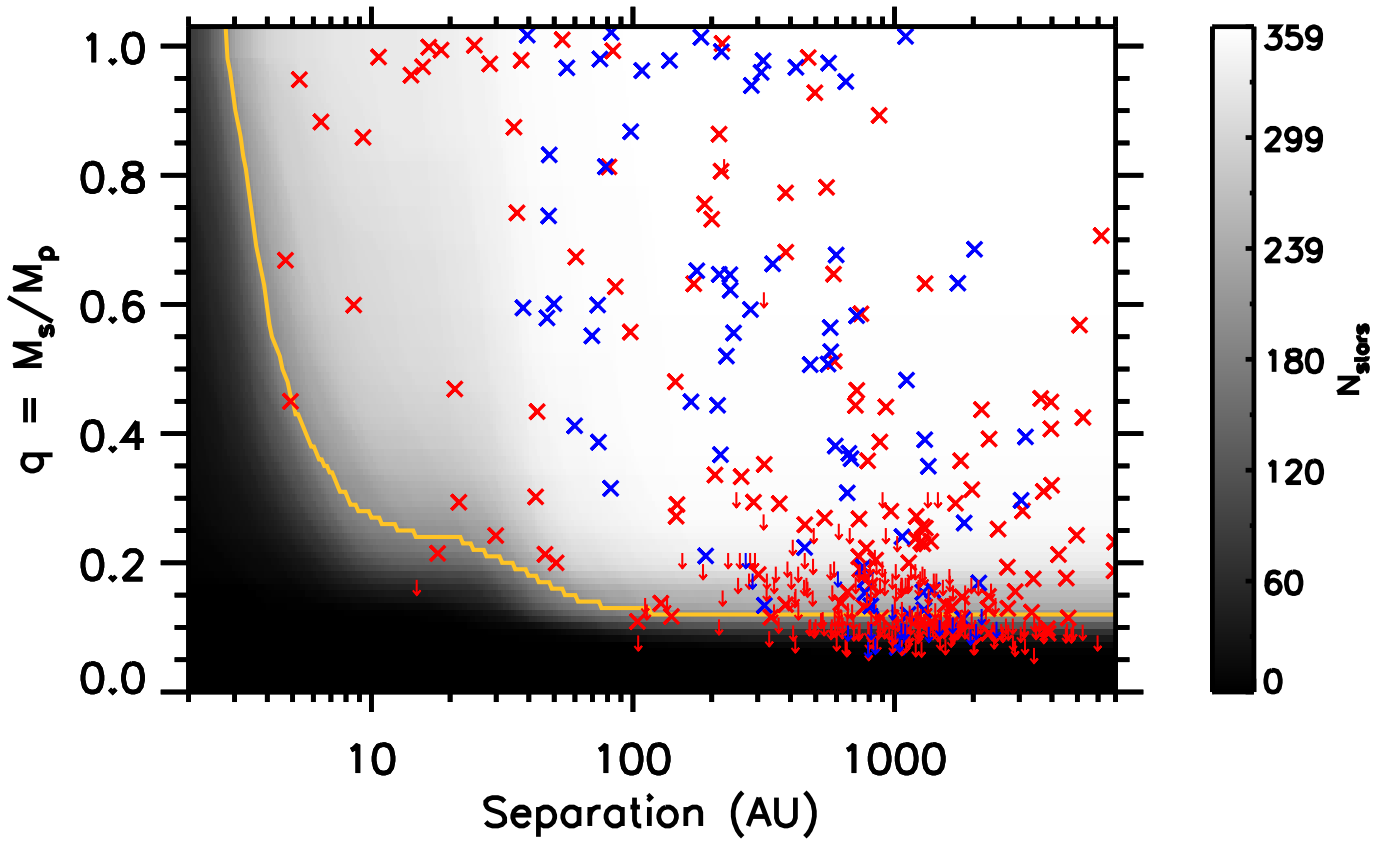}
 \caption{Detections and detection limits for our survey, in terms of binary mass ratio $q = m_s/m_p$ (if it were a bound companion) and physical projected separation $\rho$ in AU. The detections are shown with red crosses for newly detected sources and blue crosses for previously identified sources, while the detection limits are shown in a shaded background trending from black (no observations sensitive to that combination of contrast and separation) to white (all 359 planet hosts that are sensitive to that combination). The orange solid line shows the median limit for the survey. None of our sample members had a previously known companion with projected physical separation $\rho < 40$ AU, demonstrating the need for NRM to probe the deep paucity of binary companions on solar-system scales.}
  \end{figure*}

 \begin{figure*}
 \epsscale{1.2}
 \plotone{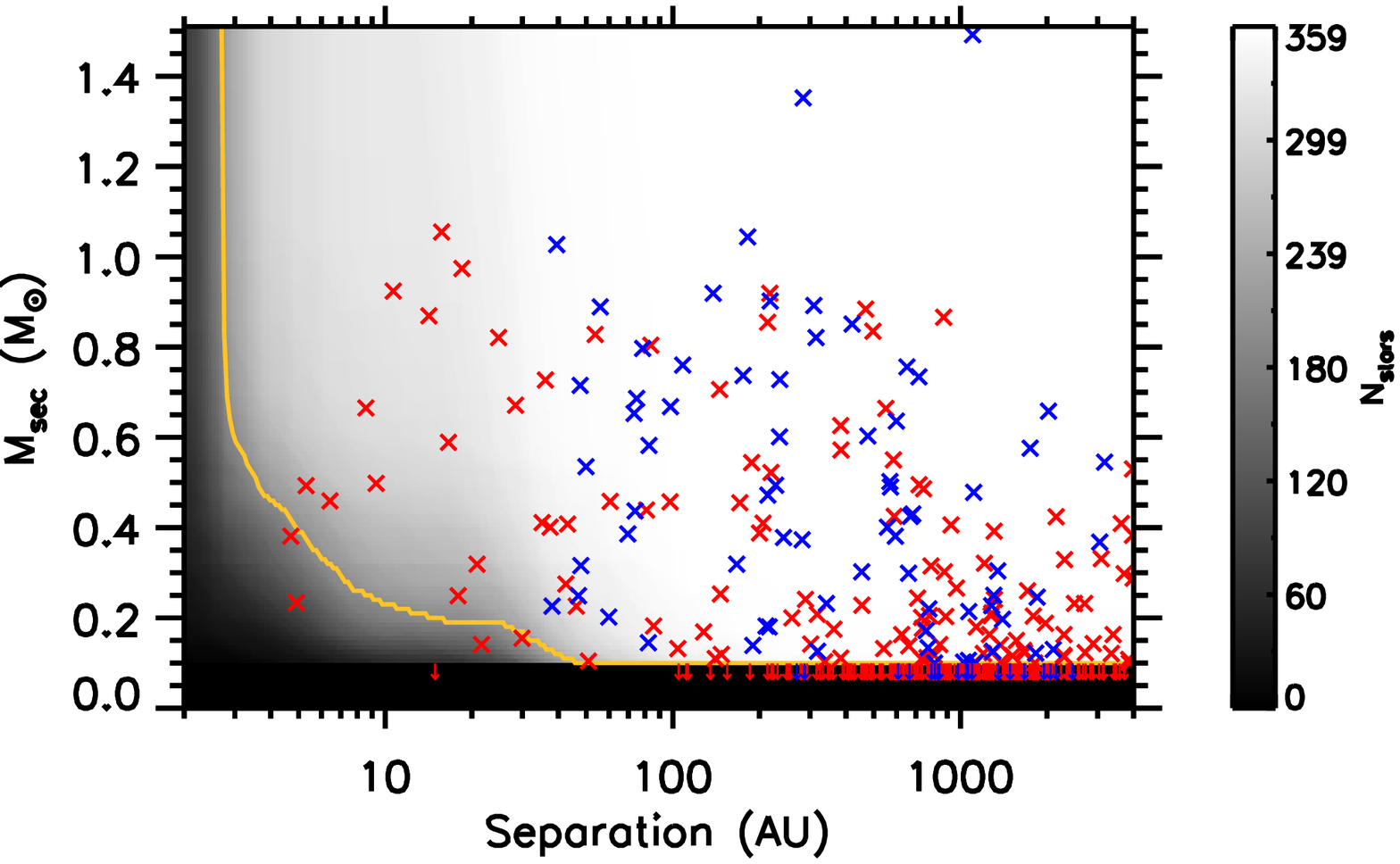}
 \caption{Detections and detection limits for our survey, in terms of candidate companion mass $M$ in $M_{\odot}$ (if it were a bound binary companion) and physical projected separation $\rho$ in AU. The detections are shown with red crosses for newly detected sources and blue crosses for previously identified sources, while the detection limits are shown in a shaded background trending from black (no observations sensitive to that combination of contrast and separation) to white (all 359 planet hosts that are sensitive to that combination). The orange solid line shows the median limit for the survey.}
  \end{figure*}

\section{Results}

\subsection{Candidate Companions to Kepler KOIs}

Our NRM observations were used to identify 26 candidate companions among the 346 KOIs observed with this technique, revealing candidates at projected separations as low as 16 mas (1/3 $\lambda /D$). We summarize the detection limits and the details of the observations in Table 2, and list the candidate companions and their observed properties in Table 3. The median target had contrast limits of $\Delta K' = 4.3$ mag at $\rho = 40$ mas (0.8 $\lambda /D$) and $\Delta K' = 3.0$ mag at $\rho = 20$ mas (0.4 $\lambda /D$). The nominal limit at $\rho = 40$ mas also applies for all larger separations, though for most targets this limit was superceded by imaging at $\rho \sim$100--150 mas.

Our imaging observations have identified 486 candidate companions among the full sample of 382 KOIs. We summarize the detection limits and the details of the observations in Table 4, and list the 427 candidate companions measured with aperture photometry in Table 5. In Table 6, we list the 43 close pairs and 7 close triples for which we used our multi-PSF fitting algorithm, as well as the observed properties of the candidate companion(s) with respect to the brightest star in the system. The median target had contrast limits of $\Delta K' = 5.5$ mag at $\rho = 150$ mas (3 $\lambda /D$), typically superceding the masking limits at $\rho \ga$100 mas. At wide separations, the median limit (corresponding to the fainter majority of stars that were not observed with the coronagraph) were $\Delta K' = 8.0$ mag at $\rho > 1$\arcsec. For coronagraphic data, we achieved contrast limits as deep as $\Delta K' > 12$ mag at wide separations ($\ga$2\arcsec, in the sky- and readnoise-limited regime).

We summarize the full set of detections and detection limits for the 359 planet hosts (excluding false positives) in terms of $\Delta K'$ (in mag) versus separation (in mas) in Figure 3, and correspondingly in terms of $K'_{sec}$ (in mag) versus separation (in mas) in Figure 4. Of our detections, 5 were detected with both masking and imaging, while 27 were detected in both coronagraphic and non-coronagraphic imaging. In case of duplication, we use the companion properties from masking over those of imaging (since they typically are more precise) and those of imaging over those of coronagraphy (since the coronagraph introduces additional astrometric and photometric uncertainties). We report the redundant detections that are not used in our analysis at the bottom of each table.

We similarly show the detections and detection limits for the 359 planet hosts in terms of mass ratio $q$ ($M_s/M_p$) versus separation (in AU) in Figure 5, and in terms of $M_{sec}$ (in $M_{\odot}$) versus separation (in AU) in Figure 6. Figure 5 demonstrates the exceptional resolution offered by NRM observations. Contrasts of $\Delta K' \sim 4$ and $\Delta K' \sim 6$ correspond to approximate mass ratios of $q \sim 0.2$ and $q \sim 0.1$ respectively; given that the mass ratio distribution of binary companions is approximately flat in the solar regime, then $\sim$80\% and $\sim$90\% of bound companions should fall above these limits. As we discuss further in Section 4.2, it indeed appears that almost all objects with higher contrast are unassociated objects seen in chance alignment. In this work, we avoid the most contaminated portions of parameter space and correct the remainder on a statistical basis; future papers in this series will present second-epoch and multi-color imaging to conclusively distinguish bound companions from interlopers.

\subsection{Probability of Chance Alignments}

Given the low galactic latitude of the Kepler field, particularly at its southeast corner, then significant background star contamination is to be expected. We have estimated this contamination in our sample using star count models that we first described in \citet{Kraus:2012bh}, which are based on those of \citet{Bahcall:1980lr} with an update to operate in the $K'$ filter. This formalism considers the Milky Way in terms of three components (thin disk, thick disk, and halo), and then integrates the 3D density distribution, convolved with the field present-day luminosity function, along each sightline to estimate the number of unassociated Milky Way stars per unit area. 

These models are sensitive to galactic structure at low galactic latitudes, but this typically takes the form of a multiplicative factor for the number of sources as a function of magnitude. We compared our model to source counts available in 2MASS, and found that the predicted counts were a factor of $\sim$2 too low, so we have scaled our source count estimates to compensate. Our predictions are similar to those from the TRILEGAL survey \citep{Girardi:2005kq}, which also requires a similar rescaling factor.

When these densities are summed for the entire target list, we find that at projected separations of $\rho < 3$\arcsec, there should be 7.1 contaminants with $K' \le 16$ and 25.4 contaminants with $K' \le 19$. In Figure 4, we show the expected contamination rate with green dashed contours drawn such that 1, 3, 10, 30, or 100 background stars would fall outside (i.e., leftward and above) the contour. The conclusion is that nearly all candidate companions with $K' > 16$ are likely to be background interlopers, but nearly all brighter companions are likely to be bound companions. Given the typical brightness and distance of our observed targets, contamination should be negligible for $\rho < 1500$ AU and $q > 0.4$.

\subsection{Comparison to Past Surveys}

We have recovered a total of 93 candidate companions that were previously reported by imaging surveys in the literature, in addition to the 413 candidates that are newly reported here. There were no candidate companions that should have fallen above our detection limits and on the NIRC2 detector that we did not recover. We mark the overlapping targets in Tables 5 and 6. The vast majority of these recoveries can be attributed to three survey programs: 46 reported by \citet{Adams:2012zr,Adams:2013ve} and \citet{Dressing:2014qf}, 16 reported by \citet{Lillo-Box:2012uq,Lillo-Box:2014ul}, and 23 reported by \citet{Law:2013fk}. Most of these recovered candidates are among the brighter and wider candidates of our sample, since past surveys were typically conducted with smaller-aperture telescopes that did not achieve the full resolution possible with NRM or the full depth possible with a 10m telescope. Every candidate companion that we detected within $\rho < 100$ mas is a new detection.

For cases with significant overlap of well-resolved companions, we can compare our angular separations to those of past surveys, testing the relative platescale calibrations of NIRC2 (e.g., Yelda et al. 2010) and the other cameras. We found that for 42 candidate companions observed by \citet{Adams:2012zr,Adams:2013ve} and \citet{Dressing:2014qf} that have projected separations of $\rho > 0.5$\arcsec, our projected separations are systematically $+3.1 \pm 0.4$\% larger, suggesting that the relative platescales should be scaled by that amount before conducting direct comparisons. Similarly, for 16 overlapping candidates observed by \citet{Lillo-Box:2012uq,Lillo-Box:2014ul} with $\rho > 0.5$\arcsec, we find our projected separations are of $+0.54 \pm 0.42$\% larger, and for 19 overlapping candidates observed by \citet{Law:2013fk} with $\rho > 0.25$\arcsec, we find our projected separations are $-1.2 \pm 0.5$\% smaller. 

After applying these platescale offsets, we find that the RMS scatter in the difference of the projected separations is $\sigma_{\rho} = 60$ mas, $\sigma_{\rho} = 50$ mas, and $\sigma_{\rho} = 27$ mas, respectively. This scatter is broadly consistent with the proper motion of these KOIs and the difference in observational epoch, and hence might be additional evidence that wider candidates are significantly contaminated by background stars (e.g., Section 4.2).

The position angles also can be similarly tested against each other. We find that the PAs reported by \citet{Adams:2012zr,Adams:2013ve} and \citet{Dressing:2014qf} are best fit to our own with a rotation of $+0.8 \pm 0.4$ degrees, with an RMS scatter of 2.8 degrees. The offset for \citet{Lillo-Box:2012uq,Lillo-Box:2014ul} is $-0.44 \pm 0.15$ degrees, with an RMS of 0.59 degrees, while the offset for \citet{Law:2013fk} is $-1.6 \pm 0.4$ degrees, with an RMS of 1.6 degrees.

The contrast measurements of \citet{Lillo-Box:2012uq,Lillo-Box:2014ul} and \citet{Law:2013fk} can't be directly tested since the measurements were conducted in the optical. However, we find that our contrast measurements agree with the $K_s$ or $K'$ measurements of \citet{Adams:2012zr,Adams:2013ve} and \citet{Dressing:2014qf} with an offset of $+0.09 \pm 0.03$ mag, with an RMS scatter of 0.20 mag on the difference. Most of the largest outliers were near the detection limits, and hence the measurements are broadly consistent within the mutual uncertainties.

A small number of other candidates were also reported by other observing programs, albeit not in sufficient numbers to directly compare our results to theirs. The optical speckle imaging programs of \citet{Howell:2011qy,Horch:2012bh,Everett:2015qd} reported 9 candidate companions in common with our program. \citet{Wang:2014uq} reported 12 candidate companions in common with our survey based on analysis of a wide range of observations downloaded from the Kepler Community Follow-Up Observing Project (CFOP), but given the heterogeneous data sources, no single calibration applies to the full dataset. Finally, \citet{Gilliland:2015qr} reported HST imaging observation with three overlapping candidates, including all components in the close triple KOI-2626.

Our sample partially overlaps with the set of KOIs that have multi-epoch RV observations that could also identify binary companions (e.g., Wang et al. 2015b). That team identified KOI-0005 to have a parabolic linear trend (Wang et al. 2014b) that they later attributed to a substellar companion (Wang et al. 2015a). Our survey confirms that there is another component in the system, but it is actually an equal-brightness companion at $\rho \sim 15.7$ AU. The observed RV trend for KOI-0005 most likely tracks the flux-weighted velocity centroid of the spectrally-unresolved pair, an effect that should be common in flux-limited samples that are subject to Malmquist bias. Their RV monitoring observations (with 5 epochs spanning one observing season) did not detect the equal-brightness companion to KOI-0289 ($\rho \sim 10.7$ AU), likely because RV changes in the flux-weighted centroid velocity are again heavily diluted. Conversely, we did not detect any companions to the system KOI-0069, for which they identified an RV trend of $12.2 \pm 0.2$ m s$^{-1}$ yr$^{-1}$. The source of the trend is likely either a substellar companion or a short-period equal-brightness companion with flux dilution. We therefore conclude that while both approaches can find close-in companions, the effects of flux dilution and Malmquist bias likely need to be factored into future analyses of RV data, as adding KOI-0005 and KOI-0069 to the previous analyses would significantly increase the inferred close-in companion fraction.

Finally, a number of candidate binaries were also identified by \citet{Kolbl:2015rc} based on spectral decomposition of Keck/HIRES optical echelle spectra. Twelve of their candidate binaries were observed in the course of our survey so far. Seven of these targets (KOI-0005, 0652, 1361, 1613, 2059, 2311, and 2813) do indeed have candidate companions within $\la$1\arcsec, though in general their predicted temperatures and contrasts do not match ours. This disagreement also has been noted by Teske et al. (2015), and could be a systematic differences in spectral information content as a function of temperature (e.g., Gullikson et al. 2016). The other five candidate binaries (KOI-0969, 2867, 3506, 3528, and 3782) do not have any imaged candidate companions that could be counterparts in our own observations, even down to projected separations of $\rho = $20--30 mas. Far less than half of our sample has a similar-brightness companion within $\rho \la 1$\arcsec, suggesting that their method is indeed able to identify some binaries. Furthermore, several of the unrecovered companions are to KOIs that are likely false positives, suggesting that these could be short-period spectroscopic binaries where Kolbl et al. detected the same (stellar) companion that is producing the transit signal (and that we could not detect). Multi-epoch RV analysis would be needed to confirm if this is the case.

\section{The Binary Population among Kepler Planet Hosts}

 \begin{figure*}
\includegraphics[scale=0.48,trim=0cm 0cm 0cm 0cm]{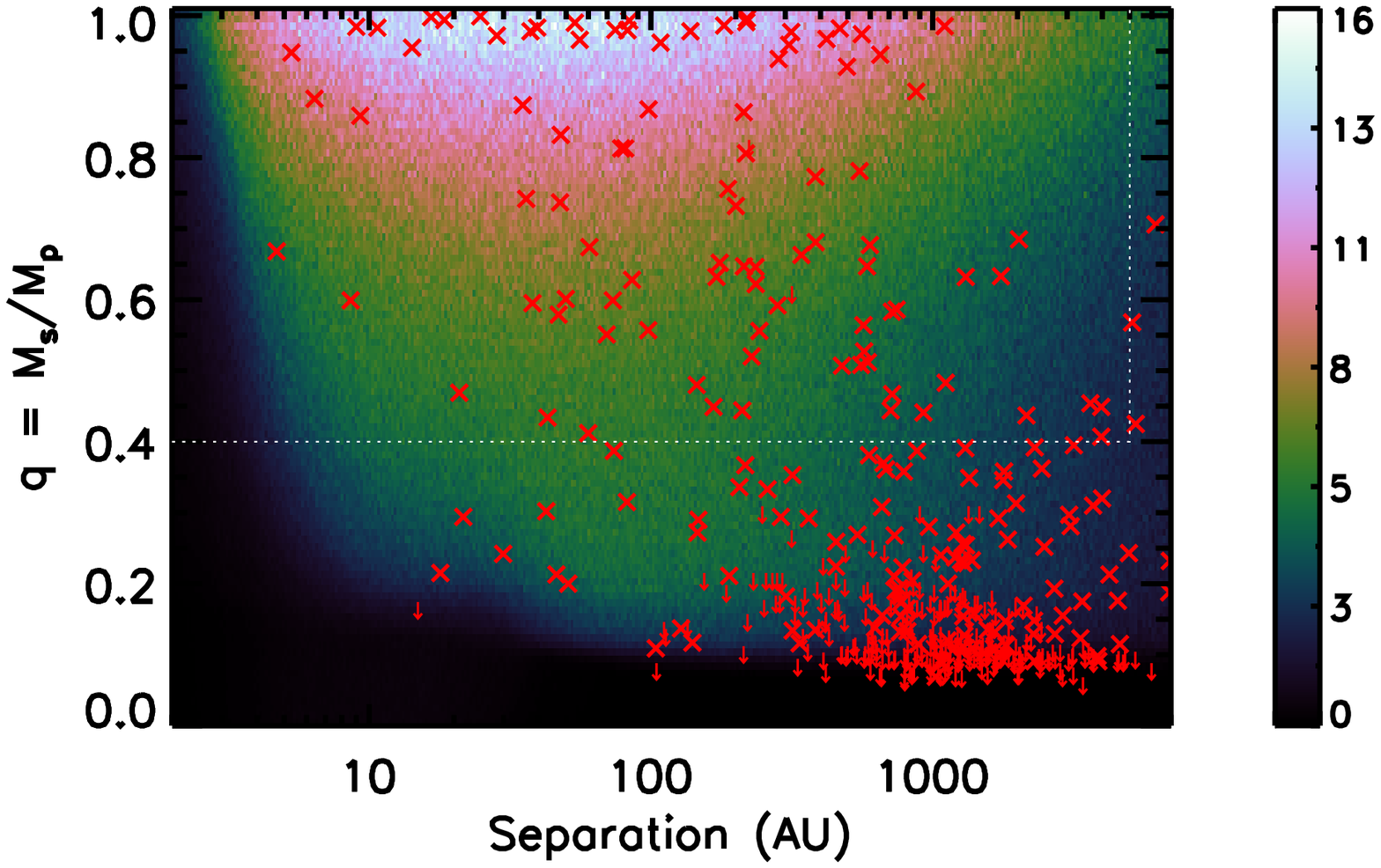}
\includegraphics[scale=0.48,trim=2cm 0cm 0cm 0cm]{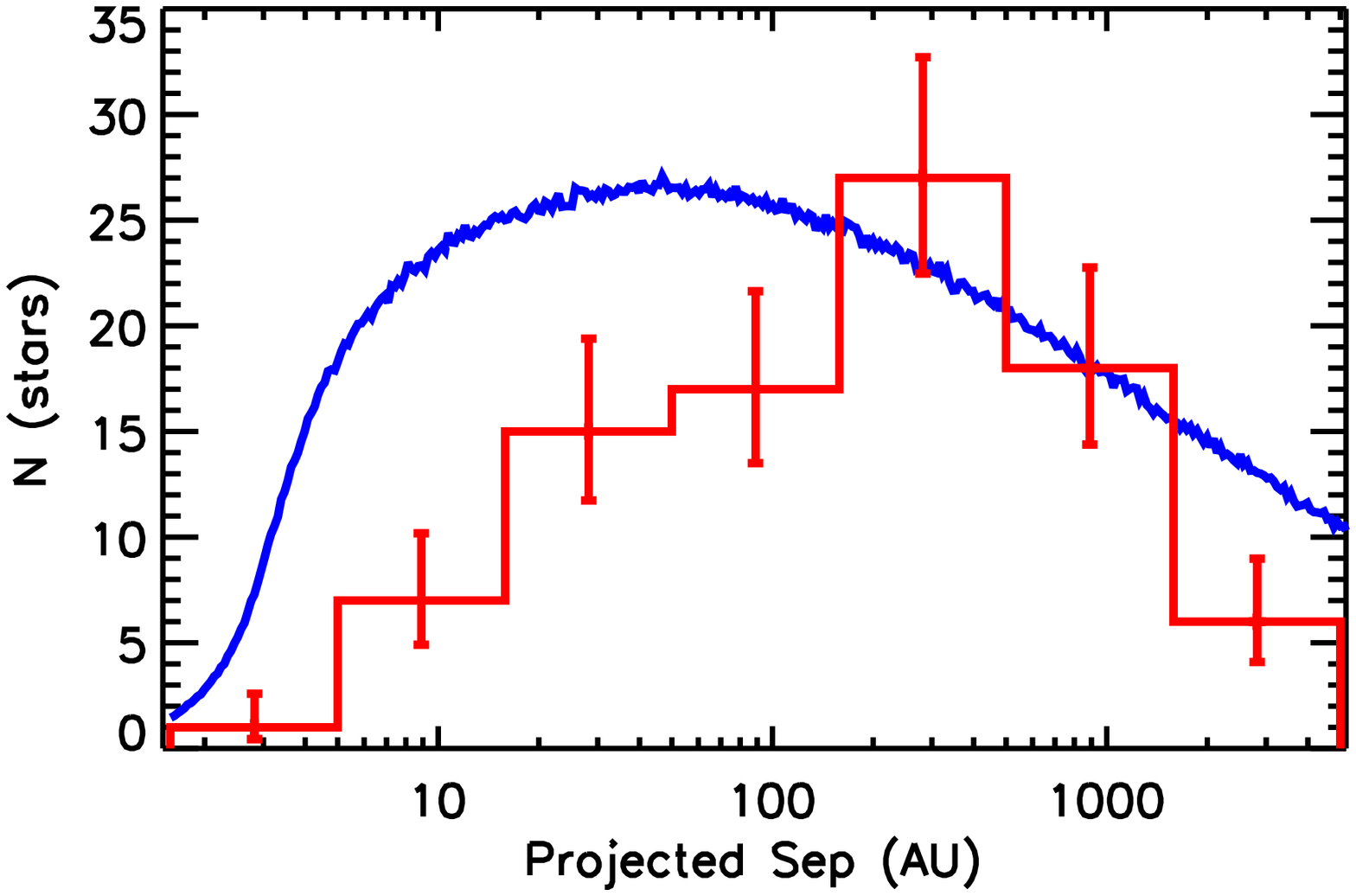}
 \caption{Left: Candidate companions (red crosses) among our sample, plotted on top of the expected density of binary companions in the observed parameter space $N(\rho,q)$ if binary companions were drawn out of the distribution reported by \citet{Raghavan:2010sz}, simulated with a random orbital phase, and then subjected to Malmquist bias and our observational detection limits. There is a clear deficit of candidates at small projected separation (denoting a paucity of short-period binaries) and an excess of faint, wide candidates (denoting the regime where background star contamination dominates). The uncontaminated space where we conduct statistical tests ($q > 0.4$, $a < 5000$ AU) is outlined with a white dotted line. Right: The marginalized distribution of projected separations, $N(\rho)$, for all companions with $q > 0.4$ (which omits nearly all background stars). The red histogram shows our observed sample, while the blue curve shows the predicted population if binary companions were drawn out of the distribution reported by \citet{Raghavan:2010sz}. As in the left panel, the deficit of close binaries is clearly evident; the distributions differ with $\chi^2 = 74.14$ or $\chi_{\nu}^2 = 10.6$ with 7 degrees of freedom (since this is a pure comparison with no fit parameters). We observe 23 companions with $\rho < 50$ AU, while the distributions of Raghavan et al. (2010) predict $58.0 \pm 7.6$ such companions; we therefore see a $4.6 \sigma$ deficit in this regime, and many of these detections likely are wide-orbiting companions that we see close only in projection. This deficit demonstrates that close binaries host planets at a lower occurrence rate than single stars or wider binary systems.}
  \end{figure*}
  
 \begin{figure*}
 \includegraphics[scale=0.9,trim=0cm 0cm 0cm 0cm]{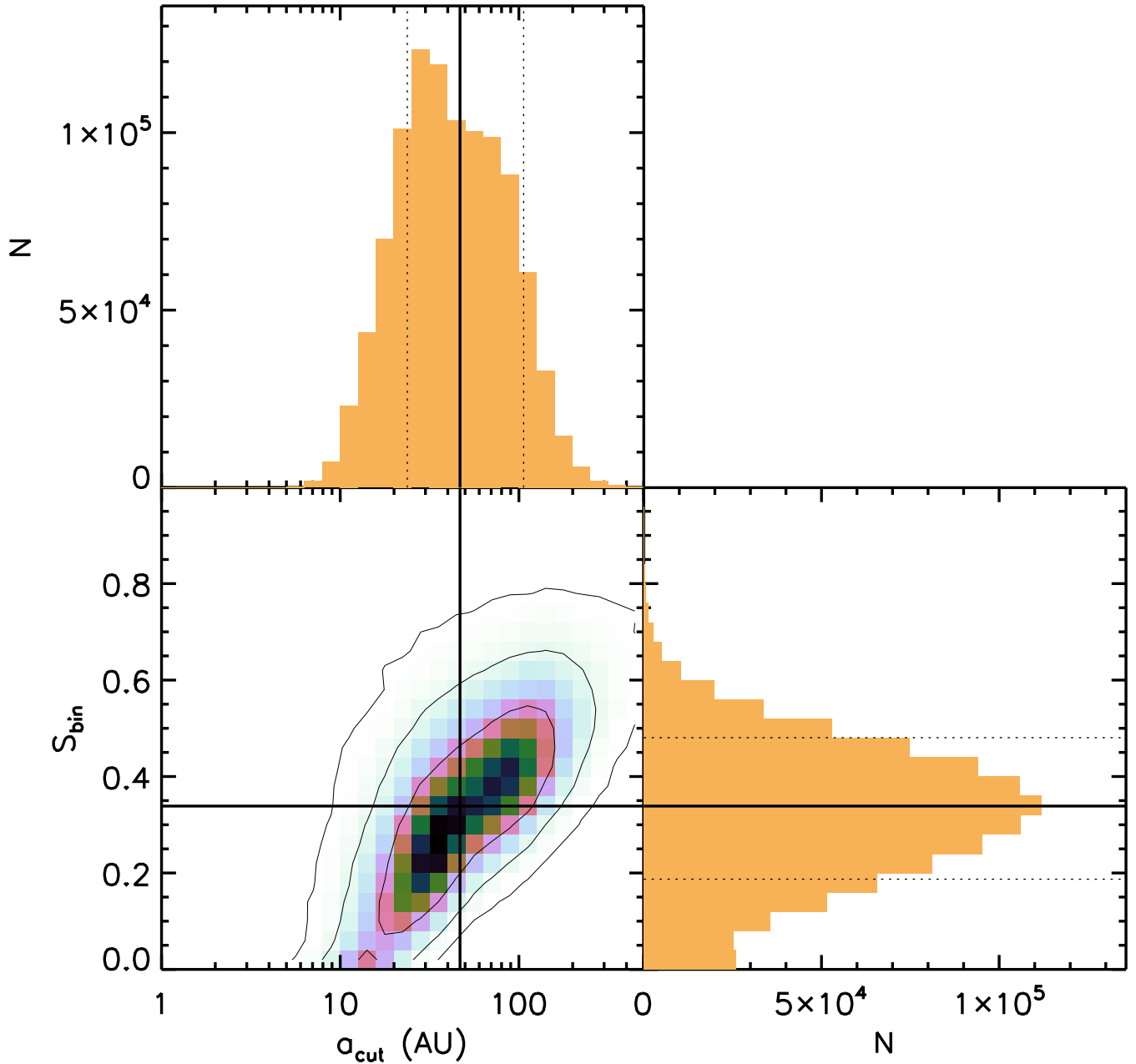}
 \caption{Joint confidence interval for the model parameters $a_{cut}$ and $S_{bin}$, shown with contours at 1$\sigma$, 2$\sigma$, and 3$\sigma$, as well as marginalized 1D histograms for the posterior of each parameter. We show the median value for each parameter, as well as the central 68\% credible interval, using solid and dashed lines respectively. The joint constraint on the two parameters is correlated, such that a larger value of $a_{cut}$ (suppressing planet occurrence in wider binaries) allows for a higher value of $S_{bin}$ (weakening the suppression). However, values of $a_{cut} \la10$ AU or $S_{bin} \ga 0.65$ are disallowed at 2$\sigma$ for any value of the other parameter. The null hypothesis ($a_{cut} = 0$ AU or $S_{bin} = 1.0$) is ruled out at $4.6\sigma$ confidence.}
  \end{figure*}
  
 \begin{figure*}
\includegraphics[scale=0.48,trim=0cm 0cm 0cm 0cm]{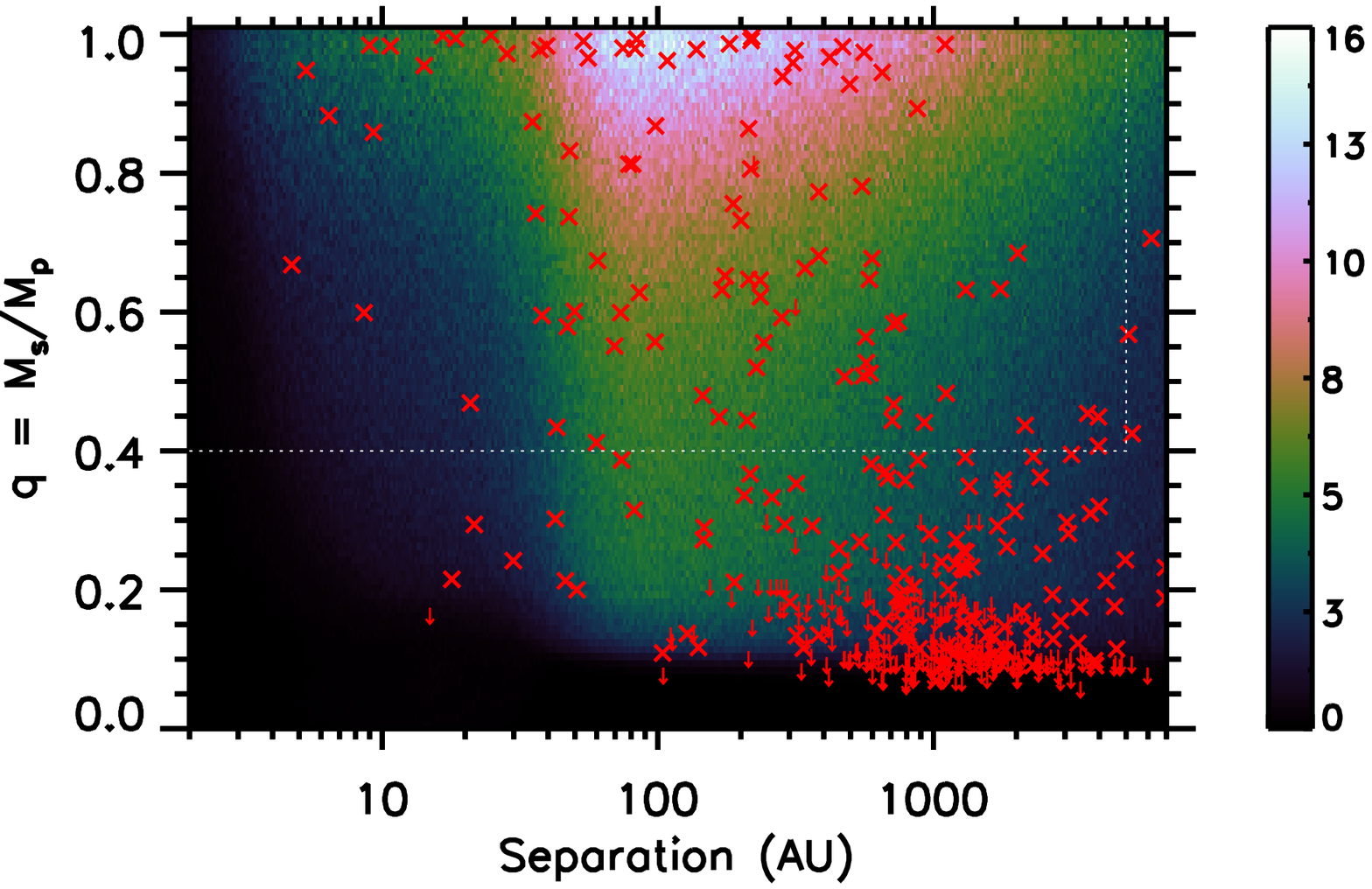}
\includegraphics[scale=0.48,trim=2cm 0cm 0cm 0cm]{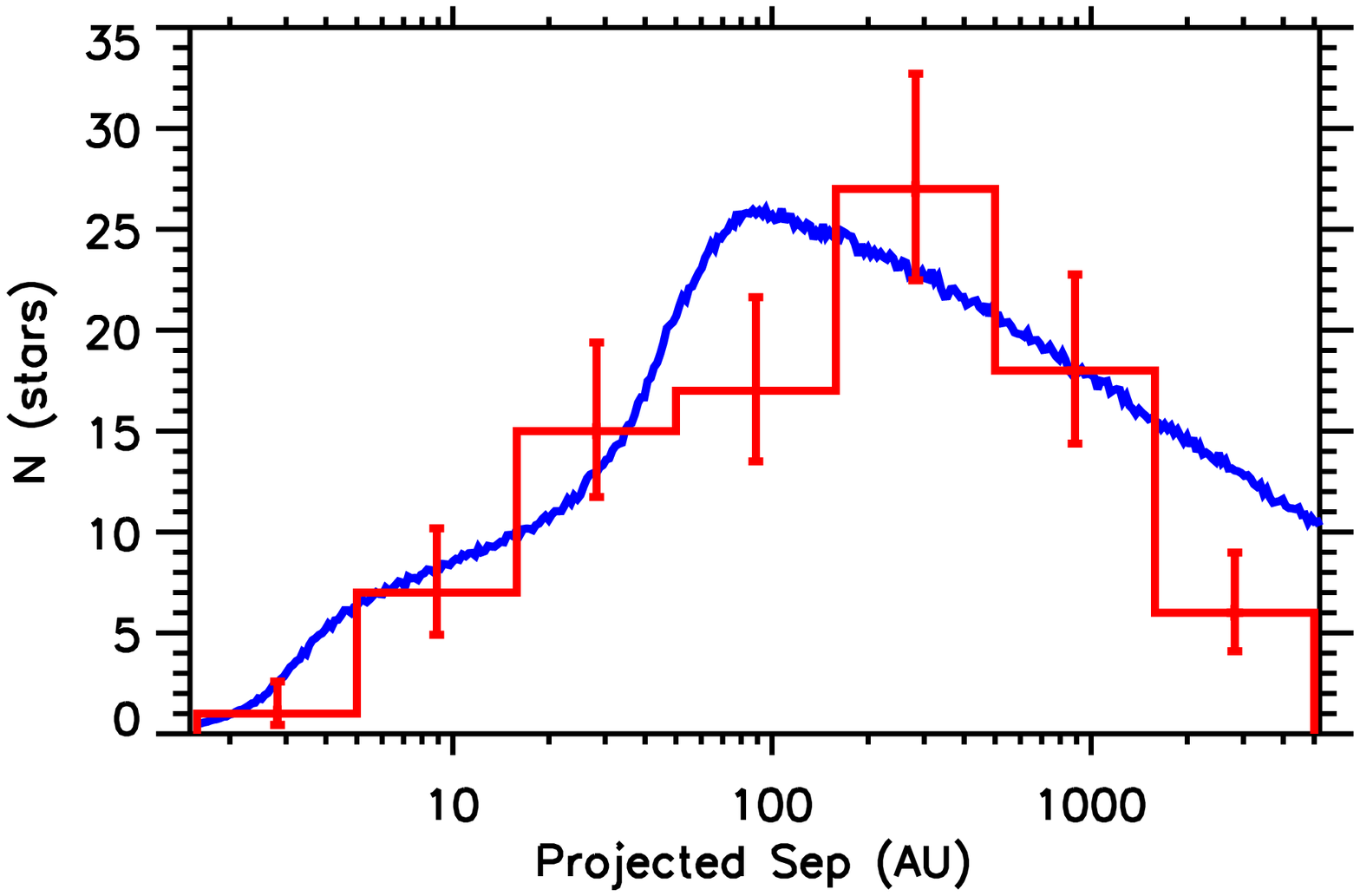}
 \caption{As in Figure 7, but for our best-fitting model that suppresses binary occurrence inside $a_{cut} = 47$ AU by a factor of $S_{bin}=0.34$. The distributions differ with $\chi^2 = 6.03$ or $\chi_{\nu}^2 = 1.21$. Given that $\chi_{\nu}^2 \sim 1$, it appears that the current data can not support more complicated models (such as with multiple cutoffs or a gradual transition). These results demonstrate the ruinous impact of close binary companions on planetary systems; these binary systems must exist, they simply don't have planets and therefore are almost totally absent from the KOI sample.}
  \end{figure*}

Binary companions should raise many dynamical barriers to the formation and survival of planetary systems, including the tidal opening of wide disk gaps, dynamical stirring of planetesimal populations, accelerated disk clearing, and ejection of planets due to long-term secular evolution of their orbits. However, our discovery of so many binary companions suggests that planet formation is possible for some binary systems, at least within restricted ranges of parameter space. The most influential feature should be the semimajor axis of the binary; wide binaries exert a weaker force on close-in planetary systems. The scale at which this transition occurs provides a strong constraint on the sum of the processes that impede planet formation and survival.

Our survey is only conducted at a single epoch, and most of our binary systems have orbital periods that are too long to measure, so we only know the instantaneous projected angular separation to each the binary companions in our sample. Due to projection effects, detection limits, and Malmquist bias, the distribution of projected separations can not be directly compared to the semimajor axis distribution of the full field population. We therefore have used a Monte Carlo routine to forward-model the binary frequency and the predicted distributions of semimajor axis, eccentricity, mass ratio, and geometric viewing angles into our observed parameter spaces. We base this model on the log-normal semimajor axis distribution and the linear-flat mass ratio and eccentricity distributions reported by \citet{Raghavan:2010sz} for solar-type stars. These distributions appear to be valid for $0.5 < M < 1.5 M_{\odot}$ stars, encompassing 90\% of our sample. The small number of targets with $M < 0.5 M_{\odot}$ might have fewer companions at $\ga$500 AU (e.g., \citealt{Reid:2001wb,Burgasser:2003mw}), but the companion frequency per dex of semimajor axis is remarkably constant in the $\la$500 AU regime across the mass range our entire sample \citep{Duchene:2013fr}.

In our forward-modeling Monte Carlo, we randomly draw binary parameters ($a$, $e$, and $q$), and then generate a random viewing angle in order to calculate the projected separation $\rho$ for that simulated binary. Finally, we multiply that detection by the binary frequency of $F = 56\%$, the fraction of our sample with detection limits sensitive to that combination of $\rho$ and $q$, and the fractional excess volume from which binaries of that mass ratio were selected ($V_{bin}(q)/V_{single}$), and add the result to a 2D histogram of the number of binary companions that were expected in our sample, $N(\rho,q)$. We repeat this process to create $10^7$ binaries, which we find is more than sufficient to minimize numerical errors in the resulting distributions. To more directly compare the projected separations, we end by marginalizing this distribution over the range of $q$ where background stars are not a significant contributor ($0.4 < q < 1.0$) to produce a 1D histogram of the number of binary companions expected in our sample, $N(\rho)$.

In Figure 7 (left), we show the 2D histogram of $N(\rho,q)$ that would be predicted for our KOI sample if the binary companions are drawn from the field binary population of \citet{Raghavan:2010sz}, as well as the projected separations and mass ratios of our observed binary companions.  The forward-model of Raghavan's binary population clearly captures the excess of equal-mass binaries due to Malmquist bias, as well as the overall variations in binary counts at 100-1000 AU. However, the predicted number of binary companions at $\rho \la $50 AU is clearly higher than the number we observe. In Figure 7 (right), we show the corresponding histogram of $N(\rho)$ (for $q > 0.4$) that we observe and the companion separation distribution that the Raghavan binary population would produce. This figure emphasizes the deep paucity of binary companions at small projected separations; while the Raghavan model would predict 58 binary companions with $\rho = $1.5--50 AU, we only observe 23 such companions. The goodness of fit for the right-hand panel is $\chi^2 = 74.1$ with 7 degrees of freedom (since there are no fit parameters), or $\chi_{\nu}^2 = 10.6$.

However, we would expect a few close companions just from projection effects for wide edge-on or eccentric systems, even if there were no binary companions with small semimajor axes. To quantify this paucity, we have constructed a model whereby the binary population to planet hosts is similar to the \citet{Raghavan:2010sz} distribution, except with a cutoff in semimajor axis $a_{cut}$ inside which the binary occurrence rate is multiplied by a suppressive factor $S_{bin}$. Again, since binary companions are unlikely to be strongly affected by much less massive planets, then this model actually corresponds to the suppression rate of planet occurrence in the (known) binary population with $a < a_{cut}$. We then reran the Monte Carlo for a range of possible values for $a_{cut}$ and $S_{bin}$ and computed the $\chi^2$ goodness of fit with respect to the observed projected separation distribution. The posterior was computed using an Metropolis-Hastings MCMC that explored the joint parameter space of the two parameters using 5 walkers producing chains of $N = 2 \times 10^5$ samples. We used a log-flat prior on $a_{cut}$ (matching the broadly logarithmic nature of the binary semimajor axis distribution; Raghavan et al. 2010) and a Beta prior on $S_{bin}$ (since it is a binomial parameter; Jeffreys 1939).

In Figure 8, we show the joint posterior on $a_{cut}$ and $S_{bin}$ and the corresponding marginalized posteriors for each parameter. There is clearly a degeneracy between the allowed values of $a_{cut}$ and $S_{bin}$, such that a less severe suppression factor is allowed if the cut is at large semimajor axis. However, the null hypothesis ($a_{cut} = 0$ AU or $S_{bin} = 1.0$) is ruled out at 4.6$\sigma$ or $>$99.99\% confidence, demonstrating that despite the degeneracy between the range and severity of the effect, the occurrence rate of short-period binaries is clearly suppressed. The median values and 68\% credible intervals for each marginalized parameter distribution are $a_{cut} = 47^{+59}_{-23}$ AU and $S_{bin} = 34^{+14}_{-15}$\%.

In Figure 9, we show the corresponding best-fit models of $N(\rho,q)$ and $N(\rho)$ for our observed population of binary companions to planet hosts, using the median values of $a_{cut}$ and $S_{bin}$ from the marginalized distributions shown in Figure 8. The resulting goodness of fit is $\chi^2 = 6.03$ for 5 degrees of freedom ($\chi_{\nu}^2 = 1.21$). Even this simple toy model produces an excellent fit to the data, arguing against the use of a more sophisticated model without a significantly larger dataset.

\section{Implications for Planet Formation and Survival}

The sharp suppression factor in our two-parameter model emphasizes the ruinous impact of close binary companions on planetary systems. There are no allowed values of $S_{bin}$ that are consistent at 2$\sigma$ with $a_{cut} \la 10$ AU or $S_{bin} \ga 0.65$, and those limits are only approached for extreme values of the other parameter (low $S_{bin}$ or high $a_{cut}$, respectively).  Clearly it is rare for planetary systems to form when a binary companion is present on a solar-system scale, as is consistent with the theoretical hurdles discussed in Section 1. The low planet occurrence rate has strong implications for the planet searches in the solar neighborhood; many of the nearest Sun-like stars (such as $\alpha$ Cen, 61 Cyg, and 40 Eri) have binary companions at relevant semimajor axes. Those targets must be weighed carefully when designing intensive surveys to identify the nearest Earth-like planets, and controversial discoveries like Alpha Cen Bb \citep{Dumusque:2012uq,Hatzes:2013fj} must be considered with this strong prior against planet existence in mind.

The trend that we see is consistent with previous studies of RV and ground-based transit discoveries, which found that the planet population in wide binary systems ($a \ga 100$ AU) is consistent with that seen for single stars \citep{Bonavita:2007dp,Mugrauer:2009hc}, but also that there might be a paucity of systems with $\rho \la 100$ AU \citep{Roell:2012rt} or equivalently that the typical orbital radii are wider for planet host stars \citep{Bergfors:2012qe}. Surveys of the KOIs at lower spatial resolution largely echo the former point \citep{Horch:2014pd}, but due to the large distance to these stars, most high-resolution imaging surveys have not probed solar-system scales for KOIs. \citet{Wang:2014uq,Wang:2014sp} have used RV trends (or the lack thereof) to probe the binary occurrence rate for KOIs on smaller spatial scales, estimating with 1--2$\sigma$ significance that binary occurrence could be suppressed even to $a \sim 1500$ AU, but with a larger suppression factor at $<$100 AU.

The paucity of planets in close binary systems parallels results seen in star-forming regions \citep{Cieza:2009fr,Duchene:2010cj,Kraus:2012qe}. The fast clearing of protoplanetary disks in these systems, when paired with the sharp suppression of planet occurrence, clearly indicates that planet formation is strongly inhibited by the dynamical influence of the companion. However, the disk-clearing effect only extends to $\rho \sim 50$ AU, whereas the disk occurrence rate is nearly 100\% for all wider systems. Also, some close binary systems (such as CoKu Tau/4; \citealt{Ireland:2008kx}) clearly still host disks and therefore might plausibly produce planets. \citet{Jang-Condell:2015rw} has used dynamical models of known planet-hosting binaries to show that in cases where a disk (otherwise equivalent to that around a single star) is simply truncated by the binary companion, then its mass reservoir is indeed still sufficient to form the planets that were observed. It therefore remains plausible that the suppression of planet occurrence results from some combination of dynamical heating of the planetesimals (e.g., \citealt{Quintana:2007lr,Haghighipour:2007vn,Rafikov:2015pb}, fast disk dissipation (out to $a \sim 50$ AU; \citealt{Alexander:2012la,Kraus:2012qe}) and the lower disk mass reservoir available for planet formation (out to 500 AU), as reported by \citet{Harris:2012oz}.

Finally, the question remains whether planet formation is totally suppressed over some range of semimajor axes. Our current results only restrict the population to a degenerate single-parameter family of values for $a_{cut}$ and $S_{bin}$, and values of $S_{bin} = 0$ would be allowed if $a_{cut} \sim $10--20 AU. Some theoretical models of planet formation make strong claims against the feasibility of planet formation in close binaries (e.g., Kley \& Nelson 2008; Zsom et al. 2011), suggesting that the handful of systems could form via processes like small-N dynamical capture (e.g., Mart\'i \& Beaug\'e 2012). However, close binary systems are now known to host RV-discovered gas giants (e.g., $\gamma$ Cep and HD 196885; \citealt{Hatzes:2003rt,Correia:2008ai}), multi-planet systems of transiting rocky planets (e.g., Kepler-444; Dupuy et al. 2016), and transiting circumbinary gas giants (Doyle et al. 2011, Welsh et al. 2012). The regime within which planet occurrence rates can be zero is therefore shrinking. Future orbit monitoring for close KOI binaries (e.g., Dupuy et al. 2016), when combined with the known orbits for systems like $\gamma$ Cep and HD 196885, will break the degeneracy between $a_{cut}$ and $S_{bin}$ by constraining $a$ for each binary and not simply the instantaneous projected separation $\rho$. These measurements also will cast further light on the reason that some planets survive in these dynamically harsh environments, whether as random unlikely events or because some binary configurations (such as low orbital eccentricity or mutual inclination) are less likely to inhibit planet formation.

\section{Summary}

We have reported the discovery of 506 candidate companions to 382 Kepler Objects of Interest, probing down to solar-system scales ($\rho = 1.5$--50 AU) using nonredundant aperture-mask interferometry and deep adaptive optics imaging with Keck-II/NIRC2. We super-resolve some binary systems to projected separations as tight as $\rho = 2$--3 AU, showing that planets might form in these dynamically active environments. However, the full distribution of projected separations for our planet-host sample more broadly reveals a deep paucity of binary companions at solar-system scales. For a field binary population, we should have found 58 binary companions with projected separation $\rho < 50$ AU and mass ratio $q > 0.4$; we instead only found 23 companions (a 4.6$\sigma$ deficit), many of which must be wider pairs that are only close in projection. When the binary population is parametrized with a semimajor axis cutoff $a_{cut}$ and a suppression factor inside that cutoff $S_{bin}$, we find that inside $a_{cut} = 47$ AU, the planet occurrence rate in binary systems is only $S_{bin}=0.34$ times that of wider binaries or single stars. In contrast, the occurrence rate of wider binary companions to planet-host stars is similar to that of the full field population, suggesting that no suppression occurs outside solar-system scales. Given the mean semimajor axis ($\bar{a} = 50$ AU) and the frequency ($F = 56\%$) of solar-type binaries, our results demonstrate that a fifth of all solar-type stars in the Milky Way are disallowed from hosting planetary systems due to the influence of a binary companion.

\acknowledgements

The authors thank Joshua Carter, Dan Fabrycky, Andrew Youdin, Jason Wright, Kaitlin Kratter, Sean Andrews, Phil Muirhead, Roman Rafikov, and Hannah Jang-Condell for interesting and helpful discussions over the course of this research program. We also thank the referee for providing a helpful critique of the work.

This work was supported by NASA Keck PI Data Awards to A. Kraus and T. Dupuy, administered by the NASA Exoplanet Science Institute, and by NASA XRP grant 14-XRP14\_2-0106 to A. Kraus. A. Mann was supposed by a Harlan J. Smith postdoctoral fellowship. D. Huber acknowledges support by the Australian Research Council's Discovery Projects funding scheme (project number DE140101364) and support by the National Aeronautics and Space Administration under Grant NNX14AB92G issued through the Kepler Participating Scientist Program.

Data presented herein were obtained at the W. M. Keck Observatory from telescope time allocated to the National Aeronautics and Space Administration through the agency's scientific partnership with the California Institute of Technology and the University of California. The Observatory was made possible by the generous financial support of the W. M. Keck Foundation. The authors wish to recognize and acknowledge the very significant cultural role and reverence that the summit of Mauna Kea has always had within the indigenous Hawaiian community. We are most fortunate to have the opportunity to conduct observations from this mountain.

\bibliographystyle{apj.bst}
\bibliography{ms.bbl}

\clearpage

\LongTables

\clearpage

\begin{deluxetable*}{lrrrrllrrl}
 \tabletypesize{\tiny}
 \tablewidth{0pt}
 \tablecaption{Kepler Objects of Interest}
 \tablehead{
 \colhead{Name} & \colhead{RA} & \colhead{DEC} & \colhead{$m_{Kp}$} & \colhead{$m_{Ks}$} & \colhead{$T_{eff}$} & 
 \colhead{$M_{prim}$} & \multicolumn{2}{c}{Distance (pc)\tablenotemark{a}} & \colhead{Ref} \vspace{0.05in}
\\
\colhead{} & \colhead{} & \colhead{} & \colhead{(mag)} & \colhead{(mag)} & \colhead{(K)} & \colhead{($M_{\odot}$)} & \colhead{(Uncorr.)} & \colhead{(Corr.)} 
  }
 \startdata

\multicolumn{9}{l}{\em Candidate or Confirmed Planet Hosts} \\ 
KOI-0001 & 19 07 14.033 & +49 18 59.04 & 11.34 &  9.85 &  5850$\pm$  117 & 0.950$^{+0.065}_{-0.065}$ &  195.6$^{+   9.7}_{-  11.6}$ &  206.4 & (Kepler-1; TRES-2) \\
KOI-0002 & 19 28 59.348 & +47 58 10.28 & 10.46 &  9.33 &  6350$\pm$  127 & 1.472$^{+0.071}_{-0.086}$ &  330.8$^{+  16.0}_{-  19.1}$ &  331.6 & (Kepler-2; HAT-P-7) \\
KOI-0003 & 19 50 50.244 & +48 04 51.07 &  9.17 &  7.01 &  4780$\pm$   95 & 0.810$^{+0.031}_{-0.036}$ &   38.3$^{+   1.5}_{-   1.8}$ &   38.3 & (Kepler-3; HAT-P-11) \\
KOI-0005 & 19 18 57.532 & +44 38 50.71 & 11.66 & 10.21 &  5753$\pm$  115 & 1.095$^{+0.105}_{-0.075}$ &  411.8$^{+  29.4}_{-  29.4}$ &  550.3 &  \\
KOI-0012 & 19 49 48.896 & +41 00 39.56 & 11.35 & 10.23 &  6638$\pm$  180 & 1.227$^{+0.206}_{-0.147}$ &  336.4$^{+ 119.6}_{-  42.7}$ &  341.4 & (Kepler-448) \\
KOI-0013 & 19 07 53.086 & +46 52 06.09 &  9.96 &  9.43 &  9071$\pm$  317 & 1.642$^{+0.133}_{-0.133}$ &  356.7$^{+  75.4}_{-  47.1}$ &  487.2 & (Kepler-13) \\
KOI-0041 & 19 25 32.637 & +41 59 24.97 & 11.20 &  9.77 &  5825$\pm$  116 & 1.042$^{+0.103}_{-0.074}$ &  285.2$^{+  20.9}_{-  17.4}$ &  285.2 & (Kepler-100) \\
\enddata
\tablenotetext{a}{Distances are listed both with and without the luminosity correction applied for flux contributions for companions that were unresolved in 2MASS. Targets were selected for observation using uncorrected distances, mostly the spectrophotometric distances described in Section 2, while isochronal distances that were corrected for multiplicity are used to calculate physical projected separations. Binary systems are weighted by a $1/V_{max}$ weighting in our analysis to avoid Malmquist bias.}
 \end{deluxetable*}

\clearpage

\begin{deluxetable*}{lcccrrrr}
\tabletypesize{\footnotesize}
\tablewidth{0pt}
\tablecaption{Keck/NIRC2 NRM Detection Limits}
\tablehead{
\colhead{Name} & \colhead{MJD} & \colhead{UT Date} & \colhead{$N_{frames}$} & \multicolumn{4}{c}{Limit (mag) at Sep. (mas) } 
\\
\colhead{} & \colhead{} & \colhead{} & \colhead{} &  \colhead{15} & \colhead{20} & \colhead{30} & \colhead{40--160}
}
\startdata
KOI-0001 & 56114.59 &    120706 &         6 &      2.43 &      3.16 &      4.11 &      4.46\\ 
KOI-0002 & 56882.48 &    140813 &         8 &      2.90 &      3.63 &      4.56 &      4.81\\ 
KOI-0003 & 56113.59 &    120705 &         5 &      2.49 &      3.22 &      4.17 &      4.55\\ 
KOI-0005 & 56153.45 &    120814 &         4 &      0.11 &      0.36 &      1.12 &      1.44\\ 
KOI-0012 & 56153.52 &    120814 &         6 &      1.42 &      2.15 &      3.10 &      3.37\\ 
KOI-0013 & 56511.35 &    130807 &         8 &      1.74 &      2.47 &      3.41 &      3.69\\ 
KOI-0041 & 56116.60 &    120708 &         6 &      2.55 &      3.28 &      4.23 &      4.54\\ 
KOI-0042 & 56869.36 &    140731 &         8 &      2.31 &      3.04 &      3.97 &      4.24\\
\enddata
\end{deluxetable*}

\clearpage

\begin{deluxetable*}{llcrrl}
\tabletypesize{\tiny}
\tablewidth{0pt}
\tablecaption{Keck/NIRC2 NRM Candidate Companions}
\tablehead{
\colhead{Name} & \colhead{MJD} & \colhead{$\Delta K'$ (mag)} & \colhead{$\rho$ (mas)} & \colhead{PA (deg)}
  }
\startdata
KOI-0005 & 56153.45 &   0.400$\pm$ 0.062 &  28.548$\pm$ 0.590 & 142.109$\pm$ 2.019\\ 
KOI-0214 & 56882.30 &   3.706$\pm$ 0.097 &  70.938$\pm$ 1.596 & 196.658$\pm$ 1.289\\ 
KOI-0289 & 56882.46 &   0.189$\pm$ 0.091 &  16.940$\pm$ 0.985 & 233.485$\pm$ 3.805\\ 
KOI-0291 & 56881.51 &   1.349$\pm$ 0.022 &  66.170$\pm$ 0.335 & 316.688$\pm$ 0.262\\ 
KOI-0854 & 56490.53 &   0.299$\pm$ 0.231 &  16.089$\pm$ 0.980 & 209.435$\pm$ 4.706\\ 
KOI-1316 & 56882.48 &   4.749$\pm$ 0.181 &  48.403$\pm$ 3.276 & 101.567$\pm$ 3.121\\ 
KOI-1397 & 56501.57 &   0.001$\pm$ 0.119 &  26.760$\pm$ 1.537 & 211.818$\pm$ 2.220\\ 
KOI-1613 & 56882.45 &   1.291$\pm$ 0.022 & 208.520$\pm$ 0.363 & 184.646$\pm$ 0.097\\
\enddata
\tablecomments{KOI-2032 B-C was also detected as part of a triple system in imaging data; we use the imaging results to achieve consistent astrometry and photometry across all three components, but we list the NRM detection here for completeness. KOI-1613 was previously identified as a candidate companion by Law et al. (2013).}
\end{deluxetable*}

\clearpage

 \begin{deluxetable*}{lccrrrrrrrrrrrr}
\tabletypesize{\footnotesize}
\tablewidth{0pt}
\tablecaption{Keck/NIRC2 Imaging Detection Limits}
\tablehead{
\colhead{Name} & \colhead{MJD} & \colhead{Filter +} & \colhead{$N_{fram}$} & \colhead{$t_{int}$} & 
\multicolumn{10}{c}{Limit (mag) at Projected Separation (mas) } \\
\colhead{} & \colhead{} & \colhead{Coronagraph} & \colhead{} & \colhead{} & \colhead{150} & \colhead{200} & \colhead{250} & \colhead{300} &
\colhead{400} & \colhead{500} & \colhead{700} & \colhead{1000} & \colhead{1500} & \colhead{2000}
}
\startdata
            KOI-0001 & 56114.59 &   Kp     &  1 &   10.00 &  4.4 &  5.3 &  5.4 &  5.8 &  5.8 &  5.9 &  6.1 &  6.0 &  6.1 &  6.0 \\ 
            KOI-0001 & 56114.59 &   Kp+C06 &  2 &   40.00 &  ... &  ... &  ... &  ... &  7.2 &  7.6 &  8.5 &  7.6 & 10.1 & 10.3 \\ 
            KOI-0002 & 56153.43 &   Kp     &  1 &   10.00 &  3.8 &  4.4 &  5.1 &  5.3 &  5.6 &  5.8 &  6.4 &  6.8 &  6.9 &  6.9 \\ 
            KOI-0002 & 56153.43 &   Kp+C06 &  2 &   40.00 &  ... &  ... &  ... &  ... &  7.2 &  7.2 &  8.1 &  9.2 & 10.2 & 10.3 \\ 
            KOI-0003 & 56113.59 &   Kp+C06 &  4 &   60.00 &  ... &  ... &  ... &  ... &  7.2 &  7.7 &  8.8 & 10.0 & 11.2 & 12.1 \\ 
            KOI-0005 & 56153.45 &   Kp     &  2 &   40.00 &  4.5 &  5.3 &  5.6 &  6.0 &  6.3 &  6.7 &  7.4 &  7.9 &  8.1 &  8.1 \\ 
            KOI-0012 & 56153.52 &   Kp     &  2 &   40.00 &  4.9 &  5.8 &  6.1 &  6.2 &  6.6 &  6.9 &  7.8 &  8.4 &  8.7 &  8.7 \\ 
            KOI-0013 & 56511.35 &   Kp     &  5 &  100.00 &  3.8 &  5.0 &  5.7 &  6.0 &  6.5 &  7.0 &  7.3 &  7.4 &  7.7 &  7.8 \\
\enddata
\end{deluxetable*}

\clearpage

\begin{deluxetable*}{lrcrrrrl}
\tabletypesize{\footnotesize}
\tablewidth{0pt}
\tablecaption{Keck/NIRC2 Imaging Candidate Companions}
\tablehead{
\colhead{Name} & \colhead{MJD} & \colhead{Filter +} & \colhead{$N_{frames}$} & 
\colhead{$\rho$} & \colhead{PA} & \colhead{$\Delta m$} & \colhead{Refs} \\
\colhead{} & \colhead{} & \colhead{Coronagraph} & \colhead{} & \colhead{(mas)} & \colhead{(deg)} & \colhead{(mag)}
}
\startdata
                KOI-0001 & 56114.59 &   Kp     &   1 &  1103.7 $\pm$ 1.5 &  136.803 $\pm$ 0.080 &  2.359 $\pm$0.006 & 6, 13 \\
                KOI-0001 & 56114.59 &   Kp+C06 &   2 &  5960.6 $\pm$ 2.7 &  112.205 $\pm$ 0.025 &  8.484 $\pm$0.119 &  \\
                KOI-0002 & 56153.43 &   Kp+C06 &   2 &  3093.2 $\pm$ 1.9 &  266.385 $\pm$ 0.035 &  7.525 $\pm$0.114 & 14 \\
                KOI-0002 & 56153.43 &   Kp     &   1 &  3872.1 $\pm$ 2.8 &   90.419 $\pm$ 0.041 &  5.965 $\pm$0.045 & 14 \\
                KOI-0012 & 56153.52 &   Kp     &   2 &   603.3 $\pm$ 1.5 &  345.657 $\pm$ 0.146 &  3.835 $\pm$0.007 &  \\
                KOI-0012 & 56153.52 &   Kp     &   2 &  1903.4 $\pm$ 2.7 &  320.307 $\pm$ 0.080 &  7.539 $\pm$0.043 &  \\
                KOI-0013 & 56511.35 &   Kp     &   5 &  1150.2 $\pm$ 1.5 &  280.780 $\pm$ 0.076 &  0.157 $\pm$0.005 & 1, 2, 6 \\
                KOI-0041 & 56116.60 &   Kp+C06 &   2 &  3434.0 $\pm$ 6.0 &  195.844 $\pm$ 0.100 & 10.049 $\pm$0.162 &  \\
\enddata
\tablecomments{When candidate companions were detected in both coronagraphic and non-coronagraphic imaging, we list the non-coronagraphic entry first and use it for all analysis; the coronagraphic detections are listed at the end of the table for completeness, but are not used further. References: 
1) Howell et al. (2011), 
2) Adams et al. (2012), 
3) Lillo-Box et al. (2012), 
4) Horch et al. (2012), 
5) Adams et al. (2013), 
6) Law et al. (2013), 
7) Dressing et al. (2014), 
8) Lillo-Box et al. (2014), 
9) Wang et al. (2014), 
10) Gilliland et al. (2014), 
11) Everett et al. (2015), 
12) Borucki et al. (2013),
13) Daemgen et al. (2009),
14) Narita et al. (2010).
}
\end{deluxetable*}

\clearpage

\begin{deluxetable*}{lcrccccl}
\tabletypesize{\footnotesize}
\tablewidth{0pt}
\tablecaption{Keck/NIRC2 PSF-Fitting Candidate Companions}
\tablehead{
\colhead{Name} & \colhead{MJD} & \colhead{Filter +} & \colhead{$N_{frames}$} & 
\colhead{$\rho$} & \colhead{PA} & \colhead{$\Delta m$} & \colhead{Refs} \\
\colhead{} & \colhead{} & \colhead{Coronagraph} & \colhead{(mas)} & \colhead{(deg)} & \colhead{(mag)}
}
\startdata
             KOI-0005 & 56153.45 &   Kp &   2 &  133.29 $\pm$ 3.21 &  306.710 $\pm$ 1.195 &  2.587 $\pm$0.138  & 9 \\
             KOI-0098 & 56866.50 &   Kp &  12 &  288.90 $\pm$ 1.52 &  144.925 $\pm$ 0.298 &  0.424 $\pm$0.003  & 1, 2, 4, 6 \\
             KOI-0112 & 56869.32 &   Kp &  12 &  101.01 $\pm$ 1.51 &  115.933 $\pm$ 0.851 &  1.105 $\pm$0.014  & 2 \\
             KOI-0174 & 56263.21 &   Kp &   4 &  583.55 $\pm$ 3.63 &   76.892 $\pm$ 0.149 &  2.567 $\pm$0.020  & 5, 6 \\
             KOI-0177 & 56869.51 &   Kp &  15 &  227.96 $\pm$ 1.54 &  217.915 $\pm$ 0.377 &  0.185 $\pm$0.004  & 6 \\
             KOI-0227 & 56501.50 &   Kp &   8 &  301.88 $\pm$ 1.63 &   69.117 $\pm$ 0.285 &  0.018 $\pm$0.001  & 1 \\
             KOI-0270 & 56869.27 &   Kp &  12 &  165.08 $\pm$ 1.51 &   64.647 $\pm$ 0.521 &  0.555 $\pm$0.003  & 2 \\
             KOI-0284 & 56869.45 &   Kp &   4 &  869.62 $\pm$ 2.05 &   98.162 $\pm$ 0.099 &  0.257 $\pm$0.004  & 2, 11 \\
\enddata
\tablecomments{When candidate companions were detected in both NRM observations and imaging observations, we use the NRM detection for all analysis; the imaging measurements are listed at the end of this table for completeness, but are not used further. The exceptions are KOI-0005 and KOI-0854, where the imaging and NRM detections represent separate companions in compact triple systems. References: 
1) Howell et al. (2011), 
2) Adams et al. (2012), 
3) Lillo-Box et al. (2012), 
4) Horch et al. (2012), 
5) Adams et al. (2013), 
6) Law et al. (2013), 
7) Dressing et al. (2014), 
8) Lillo-Box et al. (2014), 
9) Wang et al. (2014), 
10) Gilliland et al. (2014), 
11) Everett et al. (2015), 
12) Borucki et al. (2013)
}
\end{deluxetable*}

\clearpage

\begin{deluxetable}{lrrr}
\tabletypesize{\footnotesize}
\tablewidth{0pt}
\tablecaption{System Properties for Candidate Binary Systems}
\tablehead{
\colhead{Name} & \colhead{$M_{sec}$} & \colhead{$q$} & \colhead{Sep}\\
\colhead{} & \colhead{($M_{\odot}$)} & \colhead{($M_s/M_p$)} & \colhead{(AU)}
}
\startdata
KOI-0005 &    1.055 &    0.968 &    15.7 \\
KOI-0214 &    0.275 &    0.302 &    42.5 \\
KOI-0289 &    0.924 &    0.983 &    10.7 \\
KOI-0291 &    0.727 &    0.742 &    36.1 \\
KOI-0854 &    0.493 &    0.948 &     5.3 \\
KOI-1316 &    0.249 &    0.215 &    17.9 \\
KOI-1397 &    0.508 &    1.016 &     9.0 \\
KOI-1613 &    0.797 &    0.813 &    78.5 \\
\enddata
\end{deluxetable}

\clearpage

\end{document}